%% file: PRD_paper_B0ToD0h0_DToK0Spipi.tex
\documentclass[letter,aps,prd,twocolumn,superscriptaddress,showpacs,preprintnumbers,amsmath,amssymb,nofootinbib]{revtex4-1}

\usepackage{graphicx} % Include figure files
\usepackage{dcolumn}  % Align table columns on decimal point

\usepackage{mathtools}

\usepackage{xspace}

\usepackage{upgreek}

\input{babarsym.tex}

\graphicspath{{ps}}

\input{common_definitions}

\begin{document}

\title{ \quad\\[1.0cm] Measurement of {\boldmath{$\cos{2\beta}$}} in {\boldmath$\Bz \to \D^{(*)} h^{0}$} with {\boldmath$\D \to \KS \pip \pim$} decays by a combined time-dependent Dalitz plot analysis of \babar\ and Belle data}

\input{BABAR_BELLE_PRD_merged_author_list_final.tex}

\begin{abstract}
We report measurements of $\sin{2\beta}$ and $\cos{2\beta}$ from a time-dependent Dalitz plot analysis of \BtoDhzero with \DtoKSpipi decays, where the light unflavored and neutral hadron $h^{0}$ is a $\pi^{0}$, $\eta$, or $\omega$ meson.
The analysis is performed with a combination of the final data sets of the \babar\ and Belle experiments 
containing $471 \times 10^{6}$ and $772 \times 10^{6}$ $\B\Bb$ pairs collected at the $\Upsilon\left(4S\right)$ resonance
at the asymmetric-energy \B factories PEP-II at SLAC and KEKB at KEK, respectively.
We measure $\sin{2\beta} = 0.80 \pm 0.14  \,(\rm{stat.}) \pm 0.06 \,(\rm{syst.}) \pm 0.03 \,(\rm{model})$ and $\cos{2\beta} = 0.91 \pm 0.22  \,(\rm{stat.}) \pm 0.09 \,(\rm{syst.}) \pm 0.07 \,(\rm{model})$.
The result for the direct measurement of the angle is $\beta = \left( 22.5 \pm 4.4  \,(\rm{stat.}) \pm 1.2 \,(\rm{syst.}) \pm 0.6 \,(\rm{model}) \right)^{\circ}$.
The last quoted uncertainties are due to the composition of the \DzerotoKSpipi decay amplitude model, which is newly established by a Dalitz plot amplitude analysis of a high-statistics $e^{+}e^{-} \to c\bar{c}$ data sample as part of this analysis.
We find the first evidence for $\cos2\beta>0$ at the level of $3.7$ standard deviations.
The measurement excludes the trigonometric multifold solution $\pi/2 - \beta = (68.1 \pm 0.7)^{\circ}$ at the level of $7.3$ standard deviations and therefore resolves an ambiguity in the determination of the apex of the CKM Unitarity Triangle.
The hypothesis of $\beta = 0^{\circ}$ is ruled out at the level of $5.1$ standard deviations, and thus \CP violation is observed in \BtoDhzero decays.
\end{abstract}


\maketitle

\tighten

{\renewcommand{\thefootnote}{\fnsymbol{footnote}}}
\setcounter{footnote}{0}

\section{Introduction}
Breaking of \CP symmetry is a small physical effect with profound consequences.
\CP violation causes particles and antiparticles to behave differently~\cite{GellMann1955,LeeYang1956,Lee1974}.
Even if the effects are tiny, \CP violation provides the only possibility to assign matter and antimatter in an absolute and convention-independent way~\cite{BigiSandaCPViolation}.
As one of the Sakharov requirements~\cite{Sakharov1967} for baryogenesis, \CP violation is a key ingredient to generate the asymmetry between matter and antimatter shortly after the big bang that governs our present matter-dominated universe.
However, \CP violation in the standard model (SM) of electroweak interactions is several orders of magnitudes too small to account for the observed baryon asymmetry of the universe~\cite{Gavela1997,RiottoTrodden1997}.
This is a strong motivation to search for additional sources of \CP violation in nature.
In the SM, the origin of \CP violation is the single irreducible complex phase in the three-family Cabibbo-Kobayashi-Maskawa (CKM) quark-mixing matrix~\cite{Cabibbo,KobayashiMaskawa}.
Testing this prediction of the Kobayashi-Maskawa theory~\cite{KobayashiMaskawa} was the main objective for the construction and operation of the first-generation asymmetric-energy \B factory experiments \babar\ at SLAC (USA) and Belle at KEK (Japan).
\babar\ and Belle discovered \CP violation in the decays of neutral and charged \B mesons~\cite{CPV_observation_BaBar,CPV_observation_Belle,directCPV_BaBar,directCPV_Belle} and experimentally confirmed the theory predictions in numerous independent measurements~\cite{BFactoriesBook}.

In particular, \babar\ and Belle observed \CP violation in the interference between the direct decays of neutral \B mesons into \CP eigenstates and the decays after \Bz-\Bzb oscillations (referred to as ``mixing-induced \CP violation'')
for the ``gold plated'' decay mode\footnote{In this article the inclusion of charge-conjugated decay modes is implied unless otherwise stated.} $\Bz \to J/\psi K_{S}^{0}$ and other decays mediated by $\bar{b} \to \bar{c}c\bar{s}$ transitions~\cite{BaBar_btoccs,Belle_btoccs}.
By performing time-dependent \CP violation measurements of $\bar{b} \to \bar{c}c\bar{s}$ transitions, \babar\ and Belle precisely determined the parameter $\sin{2\beta} \equiv \sin{2\phi_1}$.\footnote{\babar\ uses the notation $\beta$ and Belle uses $\phi_1$; hereinafter $\beta$ is used.}
The angle $\beta$ of the CKM Unitarity Triangle is defined as $\arg\left[-V^{}_{cd} V^{*}_{cb} / V^{}_{td} V^{*}_{tb} \right]$, where $V_{ij}$ denotes a CKM matrix element.
The current world average measured from $\bar{b} \to \bar{c}c\bar{s}$ transitions is $\sin{2\beta} = 0.691 \pm 0.017$~\cite{HFAG}, which corresponds to an uncertainty on the angle $\beta$ of $0.7^{\circ}$.
However, inferring the \CP-violating weak phase $2\beta$ from the measurements of $\sin{2\beta}$ is associated with the trigonometric two-fold ambiguity, $2\beta$ and $\pi - 2\beta$ (a four-fold ambiguity in $\beta$),
and therefore to an ambiguity in the determination of the apex of the CKM Unitarity Triangle.
%%%%%%%%%%%%%%%%%%%%%%%%%%%%%%%%%%%%%%%%%%%%%%%%%%%%%%%%%%%%%%%%%%%%%%%%%%%%%%%%%%%%%%%%%%%%%%%%%%%%%%%%%
%%% Figure: Feynman graph B0->D_CP h0
%%%%%%%%%%%%%%%%%%%%%%%%%%%%%%%%%%%%%%%%%%%%%%%%%%%%%%%%%%%%%%%%%%%%%%%%%%%%%%%%%%%%%%%%%%%%%%%%%%%%%%%%%
\begin{figure}
\includegraphics[width=0.4\textwidth]{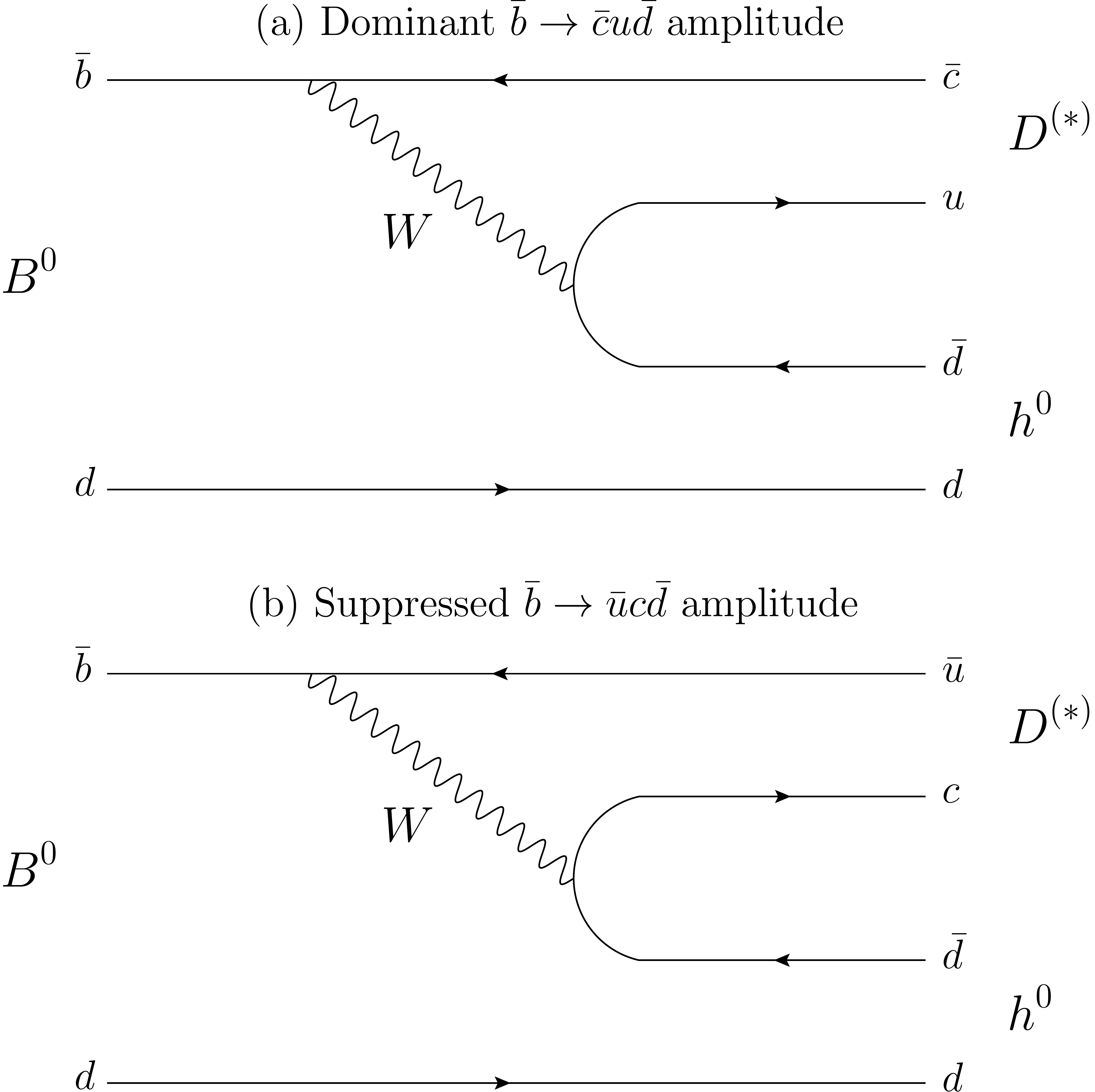}
\caption{
Feynman diagrams mediating \BtoDhzero decays:
a) the dominant $\bar{b} \to \bar{c}u\bar{d}$ tree-level amplitudes,
and b) the highly-suppressed $\bar{b} \to \bar{u}c\bar{d}$ tree-level amplitudes.
}
\label{fig:feynman_graphs_BtoDh0}
\end{figure}

The trigonometric ambiguity can be resolved experimentally by the measurements of \B meson decays that involve multibody final states.
Decay modes such as $\Bz \to J/\psi \KS \piz$~\cite{BaBar_BToJPsiK0Spi0_2005,Belle_BToJPsiK0Spi0_2005}, $\Bz \to \Dstarp \Dstarm \KS$~\cite{BaBar_BToDstarDstarK0S_2006,Belle_BToDstarDstarK0S_2007},
$\Bz \to \KS \Kp \Km$~\cite{BaBar_BToKSKK_2012,Belle_BToKSKK_2010}, $\Bz \to \KS \pip \pim$~\cite{BaBar_BToKSpipi_2009,Belle_BToKSpipi_2009},
and $\Bz \to \D^{(*)} h^{0}$ with $\D \to \KS \pip \pim$ decays (abbreviated as $\Bz \to \left[ \KS \pip \pim \right]^{(*)}_{D} h^{0}$)~\cite{BondarGershonKrokovny2005,Belle_D0h0_2006,BaBar_D0h0_2007_threebody,Belle_D0h0_2016} enable measurements of $\cos{2\beta}$ in addition to $\sin{2\beta}$.
Although $\sin{2\beta}$ is precisely measured, the experimental uncertainties on $\cos{2\beta}$ are sizable. Currently, the most precise single measurement has an uncertainty of approximately $\pm 0.36$
on the value of $\cos{2\beta}$~\cite{Belle_D0h0_2016}.
However, no previous single measurement has been sufficiently sensitive to establish the sign of $\cos{2\beta}$ that would resolve the trigonometric ambiguity without any assumptions.
The strongest constraint in the direct estimation of the angle $\beta$ was obtained by a measurement of $\Bz \to \KS \Kp \Km$ decays by \babar~\cite{BaBar_BToKSKK_2012},
which could resolve the ambiguity at the level of $4.8$ standard deviations.
However, $\Bz \to \KS \Kp \Km$ decays do not provide a theoretically clean probe for the \CP-violating weak phase $2\beta$ and only provide access to an effective weak phase $\beta_{\mathrm{eff}}$,
because at leading order $\Bz \to \KS \Kp \Km$ decays are not mediated by tree-level amplitudes but by quantum-loop (``penguin'') transitions.

An experimentally elegant and powerful approach to access $\cos{2\beta}$ and to resolve the trigonometric ambiguity is provided by \BtoDhzero with \DtoKSpipi decays~\cite{BondarGershonKrokovny2005,Belle_D0h0_2006,BaBar_D0h0_2007_threebody,Belle_D0h0_2016},
where $h^{0} \in \{\pi^{0}, \eta, \omega \}$ denotes a light unflavored and neutral hadron. The decay $\Bz \to \D^{*} \omega$ is not considered in this analysis.
As shown in Fig.~\ref{fig:feynman_graphs_BtoDh0}, the \BtoDhzero decay is mediated only by tree-level amplitudes, and to a good approximation only by color-suppressed, CKM-favored $\bar{b} \to \bar{c}u\bar{d}$ tree amplitudes.
Additional contributions from color-suppressed and doubly Cabibbo-suppressed $\bar{b} \to \bar{u}c\bar{d}$ amplitudes carry different weak phases, but are suppressed by a factor of $\lvert V^{}_{ub} V^{*}_{cd} / V^{}_{cb} V^{*}_{ud} \rvert \approx 0.02$
relative to the leading amplitudes, and can be neglected at the experimental sensitivity of the presented measurement.
The \DzerotoKSpipi decay exhibits complex interference structures that receive resonant and nonresonant contributions
from a rich variety of intermediate \CP eigenstates and quasi-flavor-specific decays to the three-body final state.
If the variations of the relative strong phase as a function of the \Dz meson three-body Dalitz plot phase space are known for \DzerotoKSpipi decays,
then both $\sin{2\beta}$ and $\cos{2\beta}$ can be measured from the time evolution of the $\Bz \to \left[ \KS \pip \pim \right]^{(*)}_{D} h^{0}$ multibody final state~\cite{BondarGershonKrokovny2005}.

In an $e^+ e^- \to \Upsilon\left(4S\right) \to \Bz\Bzb$ event, the time-dependent decay rate of the $\Bz \to \left[ \KS \pip \pim \right]^{(*)}_{D} h^{0}$ signal decays depends on the \Dz and \Dzb decay amplitudes as a function of the three-body Dalitz plot phase space
and on the \CP-violating weak phase $2\beta$, and is proportional to:
\begin{align}
  & \frac{ e^{ \frac{-\lvert \Delta t \rvert}{\tau_{\Bz} } } }{2} \Big\{ \left[ \lvert \mathcal{A}_{\Dzb} \rvert^{2} + \lvert \mathcal{A}_{\Dz} \rvert^{2} \right] \nonumber \\
  & \quad - q \left( \lvert \mathcal{A}_{\Dzb} \rvert^{2} - \lvert \mathcal{A}_{\Dz} \rvert^{2} \right) \cos(\Delta m_{d} \Delta t) \nonumber \\
  & \quad + 2 q \eta_{h^{0}} \left( -1 \right)^{L} \mathrm{Im} \left( e^{-2 i \beta} \mathcal{A}_{\Dz} \mathcal{A}_{\Dzb}^{*} \right) \sin(\Delta m_{d} \Delta t) \Big\} \mathrm{.} \label{equation:decay_rate}
\end{align}
The symbol \Deltat denotes the proper-time interval between the decays of the two \B mesons produced in the $\Upsilon(4S)$ event.
The factor $q = +1$ ($-1$) represents the $b$-flavor content when the accompanying \B meson is tagged as a \Bz(\Bzb).
The parameters $\tau_{\Bz}$ and $\Delta m_{d}$ are the neutral \B meson lifetime and the mass difference between the physical eigenstates of neutral \B mesons (``\Bz-\Bzb oscillation frequency''), respectively.
The quantity $\eta_{h^{0}} = (-1, -1, +1)$ is the \CP eigenvalue of $h^{0} = (\piz, \eta, \omega)$, and the variable $L$ is the orbital angular momentum of the $\D h^{0}$ and $\Dstar h^{0}$ system.
The relation $\eta_{h^{0}} \left( -1 \right)^{L}$ equals $-1$ for $\D h^{0}$, and $+1$ for $\Dstar h^{0}$ ($h^{0} \neq \omega$). In this analysis, we consider only $\Dstar \to \D \pi^{0}$ decays,
so an additional factor of $-1$ that should be included for $\Dstar \to \D \gamma$ decays need not be considered~\cite{BondarGershon2004}.
The \Dz and \Dzb decay amplitudes $\mathcal{A}_{\Dz} \equiv \mathcal{A}( \MsquaredKSpiRS, \MsquaredKSpiWS )$ and $\mathcal{A}_{\Dzb} \equiv \mathcal{A}( \MsquaredKSpiWS, \MsquaredKSpiRS )$
depend on the position within \DzerotoKSpipi Dalitz plot phase space defined by the Lorentz-invariant
variables $\MsquaredKSpiRS \equiv (p_{\KS} + p_{\pi^{-}})^{2}$ and $\MsquaredKSpiWS \equiv (p_{\KS} + p_{\pi^{+}})^{2}$,
where the symbol $p_i$ represents the four-momentum of a final state particle $i$.

Eq.~(\ref{equation:decay_rate}) assumes no \CP violation in \Bz-\Bzb mixing and no direct \CP violation in \BtoDhzero decays.
In our previous time-dependent \CP violation analysis combining \babar\ and Belle data~\cite{Roehrken2015}, we determined the parameter $\mathcal{C}$ that measures direct \CP violation in two independent samples of \BtoDhzero decays.
Using \D meson decays both to \CP eigenstates $D_{\CP} \to K^{+}K^{-}$, $K_{S}^{0}\pi^{0}$, and $K_{S}^{0}\omega$, and using the high-statistics control sample provided by the CKM-favored $\Dzb \to \Kp \pim$ decay mode, 
no evidence for direct \CP violation was found in either case~\cite{Roehrken2015}. This justifies the assumption of no direct \CP violation in \BtoDhzero decays for the present measurement.

The last term in Eq.~(\ref{equation:decay_rate}) can be rewritten as:
\begin{align}
  \mathrm{Im} \left( e^{-2 i \beta} \mathcal{A}_{\Dz} \mathcal{A}_{\Dzb}^{*} \right) = & \ \mathrm{Im} \left( \mathcal{A}_{\Dz} \mathcal{A}_{\Dzb}^{*} \right) \cos{2\beta} \nonumber \\
  & - \mathrm{Re} \left( \mathcal{A}_{\Dz} \mathcal{A}_{\Dzb}^{*} \right) \sin{2\beta}. \label{equation_sine_cosine}
\end{align}
Eq.~(\ref{equation_sine_cosine}) allows the measurement of $\sin{2\beta}$ and $\cos{2\beta}$ as independent parameters by a time-dependent Dalitz plot analysis of \BtoDhzero with \DtoKSpipi decays.

Although elegant and appealing, the measurements of $\sin{2\beta}$ and $\cos{2\beta}$ in \BtoDhzero with \DtoKSpipi decays are experimentally challenging and technically demanding.
The branching fractions of these \B and \D meson decays are low, at the $\mathcal{O}(10^{-4})$ and $\mathcal{O}(10^{-2})$ level, respectively.
These decay modes have neutral particles in the final states that lead to large backgrounds and low reconstruction efficiencies.
In addition, either a detailed \DzerotoKSpipi decay amplitude model or other experimental knowledge of the relative strong phase as a function of the \Dz meson three-body Dalitz plot phase space
is required as input to perform the time-dependent Dalitz plot analysis of \BtoDhzero with \DtoKSpipi decays.

Time-dependent Dalitz plot analyses of \BtoDhzero with \DtoKSpipi decays have been previously performed separately by \babar\ and Belle.
However, neither experiment was sensitive enough to establish \CP violation~\cite{Belle_D0h0_2006,BaBar_D0h0_2007_threebody,Belle_D0h0_2016}.
Some of the measurements obtained results far outside of the physical region of the parameter space~\cite{Belle_D0h0_2006},
and used different \DzerotoKSpipi decay amplitude models~\cite{Belle_D0h0_2006,BaBar_D0h0_2007_threebody}, which complicates
the comparison or the combination of the individual results.

In this article, we present measurements of $\sin{2\beta}$ and $\cos{2\beta}$ by a time-dependent Dalitz plot analysis of \BtoDhzero with \DtoKSpipi decays
that combines the \babar\ and Belle data samples, totaling $1.1\,\mathrm{ab}^{-1}$ collected at the $\Upsilon\left(4S\right)$ resonance.
In a recent combined analysis of the related decay, $\Bzb \to D^{(*)}_{\CP} h^{0}$ with $D_{\CP}$ denoting neutral \D mesons reconstructed as two-body \CP eigenstates,
we demonstrated the technical feasibility and the physical advantage of the simultaneous analysis of the data collected by the \babar\ and Belle experiments~\cite{Roehrken2015}.
In the present measurement, the benefit is two-fold:
first, the combination of the \babar\ and Belle data samples improves the achievable experimental precision by effectively doubling the statistics available for the measurement;
second, the combined approach enables common assumptions and the same \DzerotoKSpipi decay amplitude model to be applied simultaneously in the analysis of the data collected by both experiments.
The approach of combining \babar\ and Belle data enables unique experimental sensitivity beyond what would be possible by combining two independent measurements, in particular for $\cos{2\beta}$.
We derive the \DzerotoKSpipi decay amplitude model from the data by a Dalitz plot amplitude analysis of a high-statistics $e^{+}e^{-} \to c\bar{c}$ data sample.
This approach ensures full control over the construction and the propagation of uncertainties of the \DzerotoKSpipi decay amplitude model,
and thus enables further improvement of the experimental sensitivity and robustness of the measurement.

The approach of combining the existing data of the \B factory experiments \babar\ and Belle results in measurements from a data sample with an integrated luminosity of more than $1\,\mathrm{ab}^{-1}$.
Data samples of comparable size are otherwise only achievable by future heavy flavor experiments: for example, the next-generation, high-luminosity \B factory experiment Belle II~\cite{BelleIITDR},
which is expected to collect a data sample of $1\,\mathrm{ab}^{-1}$ by the year 2020 and is designed to collect $50\,\mathrm{ab}^{-1}$ by 2025.
As such, the approach of combining the data from the first-generation asymmetric-energy \B factory experiments
enables not only unique experimental precision, but also demonstrates the discovery potential of Belle II at an early phase of the experiment.

The paper is structured as follows:
Sect.~\ref{sec:detectors} introduces the \babar\ and Belle detectors and discusses the data sets used in the present analysis.
In Sect.~\ref{sec:determination_of_the_D_decay_amplitudes}, the Dalitz plot amplitude analysis to determine the \DzerotoKSpipi decay model from a high-statistics $e^{+}e^{-} \to c\bar{c}$ data sample collected by Belle is described.
Sect.~\ref{sec:Bdecay_timedependent_Dalitz_plot_analysis} presents the measurements of $\sin{2\beta}$ and $\cos{2\beta}$ by a time-dependent Dalitz plot analysis of \BtoDhzero with \DtoKSpipi decays combining the \babar\ and Belle data sets.
In Sect.~\ref{sec:Bdecay_interpretation_of_the_results}, the significance of the obtained results is studied.
Finally, Sect.~\ref{sec:conclusion} concludes the paper.
The paper is accompanied by a Letter in Physical Review Letters~\cite{accompanyingPRLpaper}.

\section{The \babar\ and Belle detectors and data sets}
\label{sec:detectors}
The results presented in this paper are based on data collected with the \babar\ detector at the PEP-II $e^+ e^-$ storage rings~\cite{PEPII}
operated at the SLAC National Accelerator Laboratory (Menlo Park, USA) and with the Belle detector at the KEKB $e^+ e^-$ storage rings~\cite{KEKB}
operated at the KEK High Energy Accelerator Research Organization (Tsukuba, Japan).
At PEP-II, $3.1~\mathrm{GeV}$ positrons collide on $9~\mathrm{GeV}$ electrons,
and at KEKB, $3.5~\mathrm{GeV}$ positrons collide on $8~\mathrm{GeV}$ electrons.
The center-of-mass (c.m.) energy of both PEP-II and KEKB is $10.58~\mathrm{GeV}$, which corresponds to the mass of the $\Upsilon(4S)$ resonance.
Due to the asymmetry of the beam energies, the $\Upsilon(4S)$ is produced with a Lorentz boost of $\beta\gamma = 0.560$ at \babar\
and $0.425$ at Belle, allowing measurement of the proper-time interval between the decays of the two \B mesons produced in $\Upsilon(4S)$ decays
from the displacement of their decay vertices. The design of \babar\ and Belle as asymmetric-energy \B factory experiments is crucial to enable time-dependent \CP violation measurements
of neutral \B mesons, as in the analysis presented in this paper.

The \babar\ and Belle detectors are large-solid-angle multipurpose magnetic spectrometers, and are described in detail elsewhere~\cite{BaBarDetector,BelleDetector,BaBarluminosity}.
The \babar\ detector consists of a five-layer, double-sided silicon vertex tracker (SVT),
a 40-layer drift chamber (DCH), an internally reflecting ring-imaging Cherenkov detector (DIRC),
and a CsI(Tl) crystal electromagnetic calorimeter (EMC) located within a super-conducting solenoid magnet that provides
a 1.5~T magnetic field. The instrumented flux return (IFR) of the solenoid magnet consists of iron plates interleaved with resistive plate chambers and, in the later runs, limited streamer tubes to detect $K_L^0$ mesons and to identify muons.

The Belle detector consists of a silicon vertex detector (SVD), a 50-layer central drift chamber (CDC), an array of aerogel threshold Cherenkov counters (ACC),
a barrel-like arrangement of time-of-flight scintillation counters (TOF), and an electromagnetic calorimeter comprised of CsI(Tl) crystals (ECL) located inside a super-conducting solenoid coil that provides a 1.5~T
magnetic field. An iron flux return located outside of the coil is instrumented to detect $K_L^0$ mesons and to identify muons (KLM). 
Two inner detector configurations were used.
A $2.0\,\mathrm{cm}$ radius beampipe and a 3-layer silicon vertex detector were used for the first sample of $152 \times 10^6 \B\Bbar$ pairs,
while a $1.5\,\mathrm{cm}$ radius beampipe, a 4-layer silicon detector, and a small-cell inner drift chamber were used to record the remaining $620 \times 10^6 \B\Bbar$ pairs~\cite{BELLESVD2}.

The Monte Carlo event generators used at \babar\ and Belle are based on EvtGen~\cite{EvtGen}, JETSET~\cite{JETSET}, and Photos~\cite{Photos}.
The \babar\ detector Monte Carlo simulation is based on Geant4~\cite{Geant4},
and the Belle detector Monte Carlo simulation is based on Geant3~\cite{Geant3}.

The first part of the analysis, described in Sect.~\ref{sec:determination_of_the_D_decay_amplitudes}, is based on a data sample of $924\,\mathrm{fb}^{-1}$ recorded at or near the $\Upsilon(4S)$ and $\Upsilon(5S)$ resonances with the Belle detector~\cite{BelleDetector}.
This data set provides a high-statistics sample of ${e^{+}e^{-} \to c\bar{c}}$ events that is used to determine the \DzerotoKSpipi decay amplitudes.
The data set provided by Belle enables a \DzerotoKSpipi yield that is about three orders
of magnitude larger than for the corresponding \B meson decay to be studied by the combined \babar+Belle approach.
Therefore, the first part of the analysis does not require the combined use of the \babar\ and Belle data sets.

The second part of the analysis, described in Sect.~\ref{sec:Bdecay_timedependent_Dalitz_plot_analysis}, is based on data samples collected at the $\Upsilon(4S)$ resonance containing
$( 471 \pm 3 )\times 10^6\, \B\Bbar$ pairs recorded with the \babar\ detector and
$( 772 \pm 11 )\times 10^6\, \B\Bbar$ pairs recorded with the Belle detector.
The combined \babar\ and Belle data set is used to perform the time-dependent Dalitz plot analysis of \BtoDhzero with \DtoKSpipi decays.

\section{Determination of the {\boldmath$\Dz \to \KS \pip \pim$} decay amplitudes by Dalitz plot amplitude analysis using Belle {\boldmath$\MakeLowercase{e^{+}e^{-} \to c\bar{c}}$} data}
\label{sec:determination_of_the_D_decay_amplitudes}

\subsection{Event reconstruction and selection}
\label{sec:reconstruction_and_selection}
The \DstarplustoDzeropisoft candidates are reconstructed from \DzerotoKSpipi decays and a low momentum (``slow'') charged pion \piplussoft.
The slow pion enables the identification of the production flavor of the neutral \D meson, which cannot be inferred directly from the self-conjugate three-body final state.
The positive (negative) charge of the \piplussoft determines the flavor of the neutral \D meson to be \Dz (\Dzb).
Neutral kaons are reconstructed in the decay mode $\KS \to \pip\pim$, with the invariant mass required to be within $15~\mevcc$ of the nominal value~\cite{PDG2016}.
Further standard requirements exploiting the displacement of the $\KS$ decay vertex from the interaction point (IP) described in Ref.~\cite{BELLEKshortselection} are applied.
For candidates reconstructed from $\Upsilon(4S)$ and $\Upsilon(5S)$ data, requirements of $p^{*}(\Dstarp) > 2.5\,\gevc$ and $p^{*}(\Dstarp) > 3.1\,\gevc$ are applied,
respectively, to reject combinatorial background and contamination from \B meson decays, where $p^{*}$ denotes the momentum in the $e^+ e^-$ c.m.\ frame.
The decay vertex of \Dstarp candidates is determined by estimating the \Dz meson production vertex from a kinematic fit. 
In the kinematic fit, the \Dz meson is constrained to originate from the $e^+ e^-$ interaction region.
The momentum resolution of soft pions is improved by a kinematic fit in which the \piplussoft is constrained to the determined \Dstarp decay vertex.

\begin{figure*}
\includegraphics[width=0.7\textwidth]{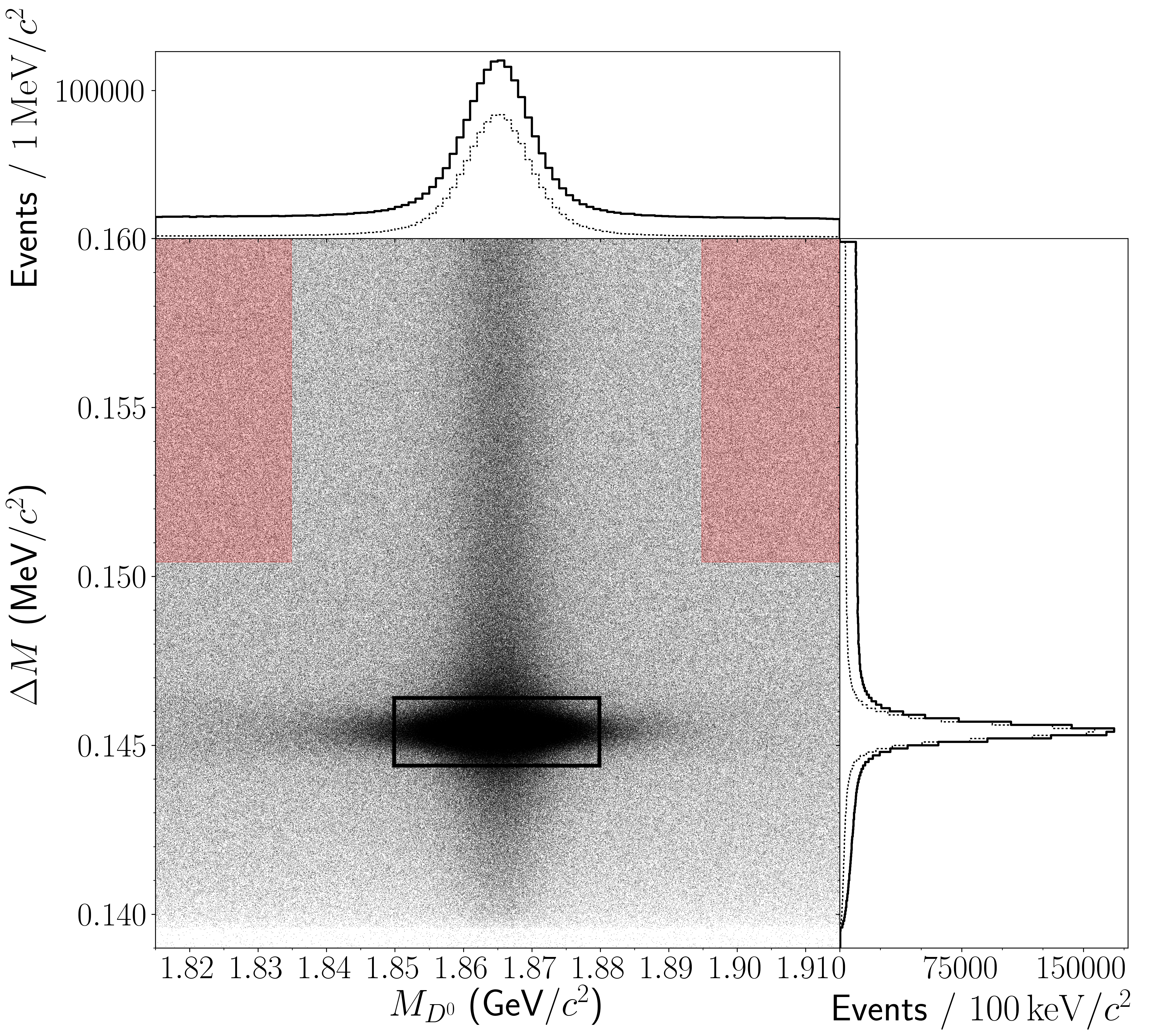}
\caption{(color online). Two-dimensional \deltaMDstarDzero and \MDzero data distributions
for \DstarplustoDzeropisoft with \DzerotoKSpipi decays reconstructed from Belle ${e^{+}e^{-} \to c\bar{c}}$ data,
and the definitions of the signal (open black rectangle) and sideband regions (filled red rectangles). The histograms on the top and at the right show one-dimensional projections for \MDzero and $\Delta M$, respectively.
In the histograms, solid lines indicate projections for one observable within the full range of the other observable,
and dashed lines represent projections in which the other observable is required to be within the signal region.
}
\label{figure:BELLE_D0mass_massdiff_fit_data}
\end{figure*}

The reconstructed charmed meson decays are characterized by two observables: the \Dz candidate mass, \MDzero, and the $D^{*+} - D^{0}$ mass difference, \deltaMDstarDzero.
Events are selected by requiring $1.825 < \MDzero < 1.905\,\gevcc$ and $140 < \deltaMDstarDzero < 150\,\mevcc$.
For the Dalitz plot fit, a narrower, signal-enhanced region is defined by requiring $(1.865 - 0.015) < \MDzero < (1.865 + 0.015)\,\gevcc$ and $(145.4 - 1.0) < \deltaMDstarDzero < (145.4 + 1.0)\,\mevcc$.
The two-dimensional \deltaMDstarDzero and \MDzero data distributions and projections of each observable are shown in Fig.~\ref{figure:BELLE_D0mass_massdiff_fit_data}.

%%%%%%%%%%%%%%%%%%%%%%%%%%%%%%%%%%%%%%%%%%%%%%%%%%%%%%%%%%%%%%%%%%%%%%%%%%%%%%%%%%%%%%%%%%%%%%%%%%%%%%%%%
%%% Figure: D0mass massdiff fit data and massdiff and D0mass plane
%%%%%%%%%%%%%%%%%%%%%%%%%%%%%%%%%%%%%%%%%%%%%%%%%%%%%%%%%%%%%%%%%%%%%%%%%%%%%%%%%%%%%%%%%%%%%%%%%%%%%%%%%

%%%%%%%%%%%%%%%%%%%%%%%%%%%%%%%%%%%%%%%%%%%%%%%%%%%%%%%%%%%%%%%%%%%%%%%%%%%%%%%%%%%%%%%%%%%%%%%%%%%%%%%%%
%%% Figure: D0mass massdiff fit data
%%%%%%%%%%%%%%%%%%%%%%%%%%%%%%%%%%%%%%%%%%%%%%%%%%%%%%%%%%%%%%%%%%%%%%%%%%%%%%%%%%%%%%%%%%%%%%%%%%%%%%%%%
\begin{figure}
\includegraphics[width=0.45\textwidth]{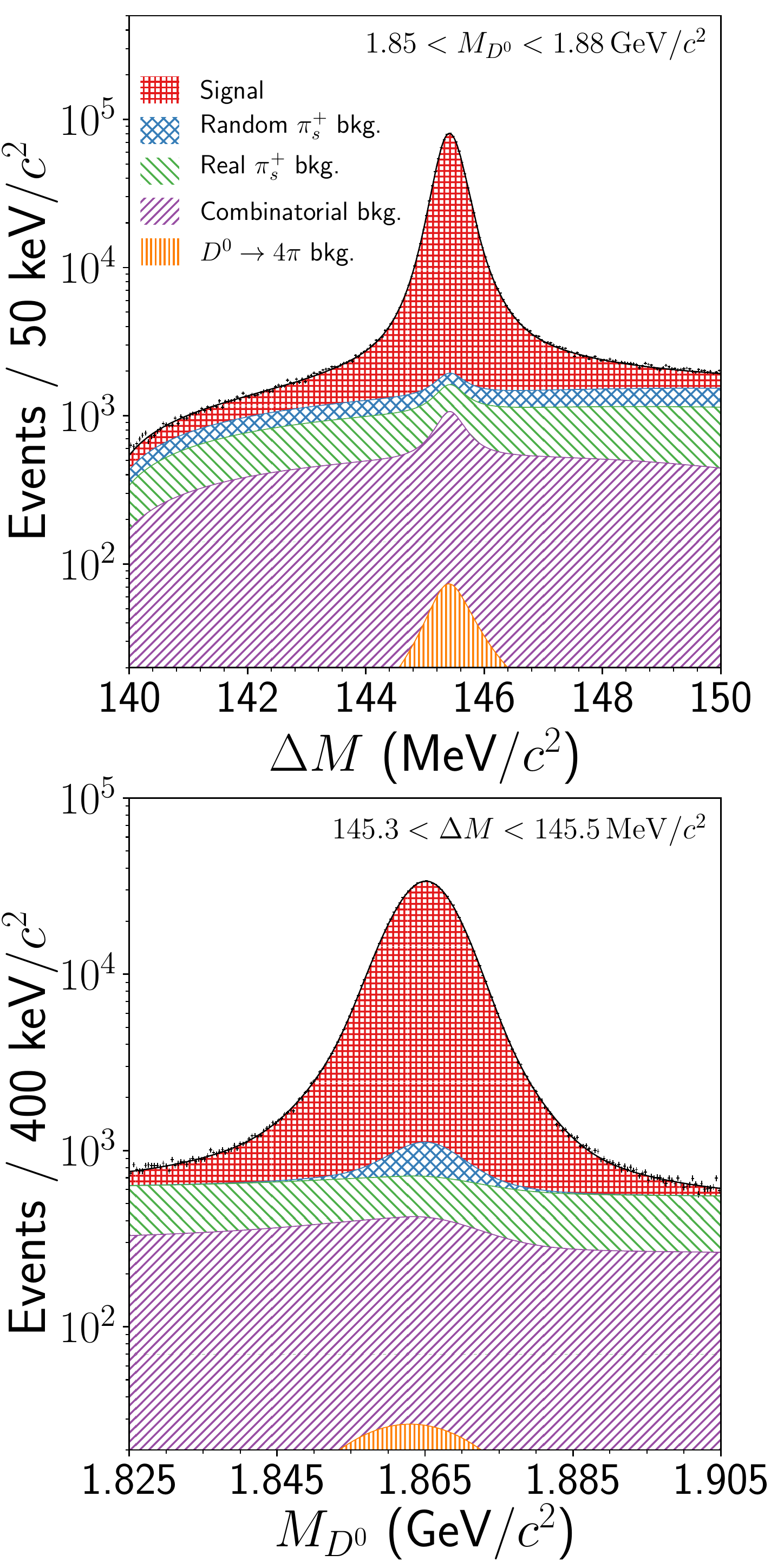}
\caption{(color online). Data distributions of \deltaMDstarDzero and \MDzero for \DstarplustoDzeropisoft with \DzerotoKSpipi decays reconstructed from Belle ${e^{+}e^{-} \to c\bar{c}}$ data (points with error bars),
and projections of the signal and background components of the fit (lines and shaded areas) as indicated in the top panel's legend.
}
\label{figure:BELLE_D0mass_massdiff_fit_projections}
\end{figure}

\subsection{Estimation of the {\boldmath$\DzerotoKSpipi$} signal and background yields}
\label{sec:determination_Dmeson_yields}
The signal and background yields are estimated by a two-dimensional unbinned maximum-likelihood (ML) fit to the \deltaMDstarDzero and \MDzero distributions.
In the fit, the shape of the \DstarplustoDzeropisoft with \DzerotoKSpipi signal decays is parameterized by the sum of four two-piece normal distributions for \MDzero and by the sum of a normal distribution,
a Johnson's SU function~\cite{JohnsonSUfunction}, a two-piece normal distribution, and a threshold function of the form $(\deltaMDstarDzero - M_{\pi^{+}})^{1/2} + a (\deltaMDstarDzero - M_{\pi^{+}})^{3/2} + b (\deltaMDstarDzero - M_{\pi^{+}})^{5/2}$ for \deltaMDstarDzero.
The width of the reconstructed \deltaMDstarDzero distribution depends on the \Dz candidate mass. The \deltaMDstarDzero distribution tends to become broader if the reconstructed \Dz mass deviates from the \MDzero peak position.
To account for this correlation, the \deltaMDstarDzero distribution is constructed by a conditional probability density function (p.d.f.) that scales the \deltaMDstarDzero width with a fourth-order polynomial function
that has the deviation of the reconstructed \MDzero from the \MDzero peak position as argument.
In the fit, the fractions and widths of the tail components relative to that of the core components are fixed to values estimated using MC simulations, and the fractions and widths of the core components are determined by the fit.

The following four separate categories are considered for the background:

The first source of background arises from the combination of correctly reconstructed \DzerotoKSpipi candidates with random tracks during reconstruction.
This ``random slow pion'' background has the same \MDzero shape as the signal, but the \deltaMDstarDzero shape follows a smooth phase space distribution that is parameterized by a threshold function.

The second background category is composed of real \piplussoft from \DstarplustoDzeropisoft decays that are combined with wrong \Dz candidates formed from random tracks or with misreconstructed real \Dz decays.
The distribution of this ``real slow pion'' background is mainly flat in \MDzero and very broad in \deltaMDstarDzero due to the reconstruction of wrong \Dz candidates,
but receives a small contribution that peaks in \deltaMDstarDzero but is broad in \MDzero due to misreconstructed real \Dz decays.
The shape of the background for wrong \Dz candidates is parameterized by a first-order polynomial function and a threshold function in \MDzero and \deltaMDstarDzero, respectively;
that for misreconstructed real \Dz decays is parameterized by a Crystal Ball function~\cite{CBfunction} and a Johnson's SU function for \MDzero and \deltaMDstarDzero, respectively.

The third background category contains background from \Dz decay modes that have the same final state as \DzerotoKSpipi decays, for example, $\Dz \to \pip\pim\pip\pim$ and $\Dz \to \KS\KS$ decays.
The $\Dz \to \pip\pim\pip\pim$ decays are effectively removed by the applied \KS selection, and $\Dz \to \KS\KS$ decays have a very small branching fraction of $\mathcal{O}(10^{-4})$.
This ``$\Dz \to 4\pi$'' background is parameterized by two Gaussian functions for \MDzero and the sum of a Gaussian function and a Johnson's SU function for \deltaMDstarDzero.
The yield of this background relative to the signal is at the sub-percent level.
The fraction of this background is fixed to the expectation value obtained from Monte Carlo (MC) simulations.

The fourth background category accounts for the remaining combinatorial background originating from random combinations of tracks.
This ``combinatorial background'' is parameterized by a first-order polynomial function in \MDzero and a threshold function in \deltaMDstarDzero.

%%%%%%%%%%%%%%%%%%%%%%%%%%%%%%%%%%%%%%%%%%%%%%%%%%%%%%%%%%%%%%%%%%%%%%%%%%%%%%%%%%%%%%%%%%%%%%%%%%%%%%%%%
%%% Table: Yields ccbar Belle
%%%%%%%%%%%%%%%%%%%%%%%%%%%%%%%%%%%%%%%%%%%%%%%%%%%%%%%%%%%%%%%%%%%%%%%%%%%%%%%%%%%%%%%%%%%%%%%%%%%%%%%%%
\begin{table*}
\caption{Signal and background yields determined by a two-dimensional fit to the \MDzero and \deltaMDstarDzero distributions of \DstarplustoDzeropisoft with \DzerotoKSpipi decays reconstructed from Belle ${e^{+}e^{-} \to c\bar{c}}$ data.}
\label{table:BELLE_yields_2D_fit_MD_DeltaM}
\begin{tabular}{l D{,}{\pm}{-1} }
\hline \hline
\multicolumn{1}{l}{Component} & \multicolumn{1}{c}{Yield} \\
\hline
\DstarplustoDzeropisoft with \DzerotoKSpipi signal & 1\,217\,300\,,\,\ 2\,000 \\
Background containing real \Dz  and random slow pions & 61\,330\,,\,\ 1\,280 \\
Background containing real slow pions and wrong \Dz & 249\,700\,,\,10\,000 \\
Background from $\Dz \to 4\pi$ & \multicolumn{1}{c}{\ \ 3\,400 (fixed)} \\
Combinatorial background & 271\,000\,,\,\ 9\,000 \\
\hline \hline
\end{tabular}
\end{table*}
In the two-dimensional fit of the \deltaMDstarDzero and \MDzero distributions, a total yield of $1\,217\,300 \pm 2\,000$ signal events is obtained.
The signal purity is $94\%$ in the signal-enhanced region used to extract the \DzerotoKSpipi decay amplitude parameters.
The results of the fit are summarized in Table~\ref{table:BELLE_yields_2D_fit_MD_DeltaM}.
The \deltaMDstarDzero and \MDzero data distributions and projections of the fit are shown in Fig.~\ref{figure:BELLE_D0mass_massdiff_fit_projections}.

\subsection{Dalitz plot amplitude analysis}
\label{sec:D_decay_Dalitz_amplitude_analysis}
\begin{figure*}
\includegraphics[width=0.99\textwidth]{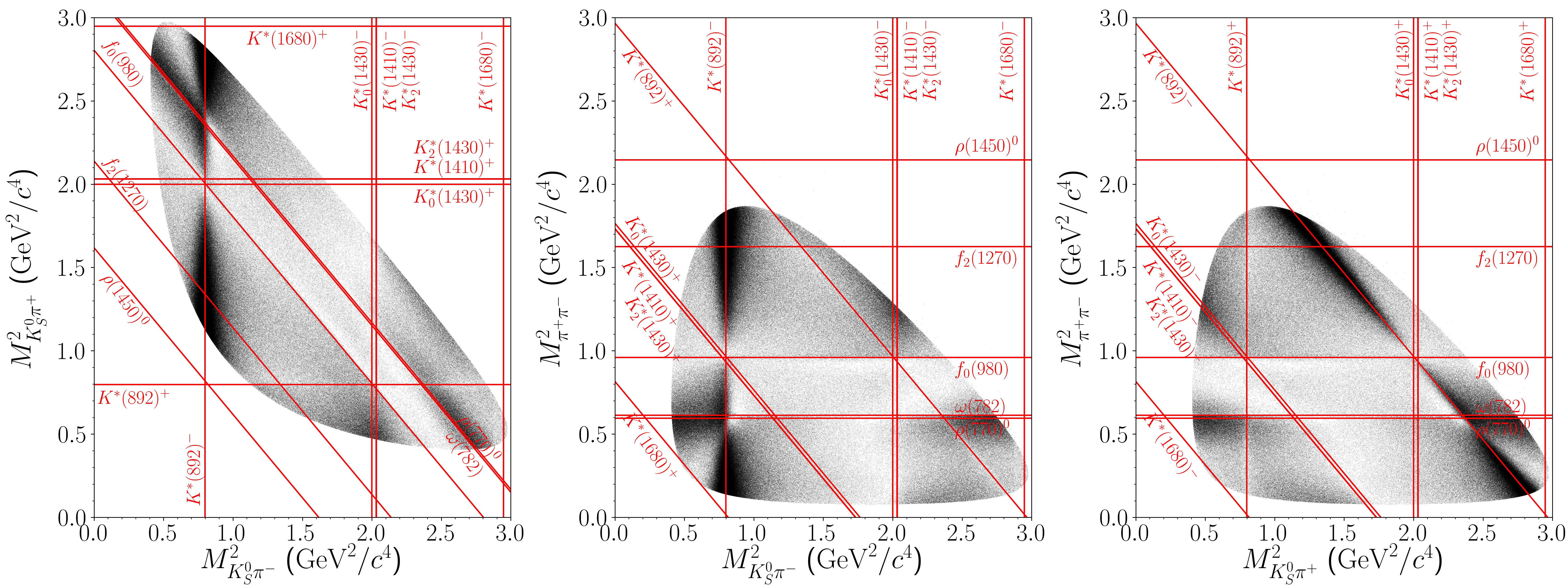}
\caption{
Dalitz plot data distributions for all three combinations of \MsquaredKSpiRS, \MsquaredKSpiWS, and \Msquaredpipi
for \DzerotoKSpipi from \DstarplustoDzeropisoft decays reconstructed from Belle ${e^{+}e^{-} \to c\bar{c}}$ data.
For illustration purposes, the approximate locations of various intermediate two-body resonances are indicated by horizontal,
vertical, and diagonal lines.
}
\label{figure_Dalitz_data_distributions}
\end{figure*}

The \DzerotoKSpipi decay proceeds via a rich variety of intermediate resonant and nonresonant modes contributing to the three-body final state.
The contributions exhibit complex interference phenomena that are observable as characteristic patterns in the three-body Dalitz plot phase space as shown in Fig.~\ref{figure_Dalitz_data_distributions}.
A Dalitz plot amplitude analysis is performed to disentangle and quantify the individual contributions.

%%%%%%%%%%%%%%%%%%%%%%%%%%%%%%%%%%%%%%%%%%%%%%%%%%%%%%%%%%%%%%%%%%%%%%%%%%%%%%%%%%%%%%%%%%%%%%%%%%%%%%%%%
%%% Figure: Dalitz data distributions
%%%%%%%%%%%%%%%%%%%%%%%%%%%%%%%%%%%%%%%%%%%%%%%%%%%%%%%%%%%%%%%%%%%%%%%%%%%%%%%%%%%%%%%%%%%%%%%%%%%%%%%%%

\subsubsection{Dalitz plot amplitude model}
\label{sec:D_decay_Dalitz_plot_amplitude_model}
To describe the resonant and nonresonant substructure and to parameterize the \DzerotoKSpipi decay amplitudes,
the isobar ansatz~\cite{ReviewDalitzPlotAnalysis} is combined with the $K$-matrix formalism~\cite{Kmatrix1995} for the $\pi\pi$ $S$-wave and the LASS parametrization~\cite{LASS} for the $K\pi$ $S$-wave.
In this approach, the \DzerotoKSpipi decay amplitudes can be written as:
\begin{widetext}
\begin{equation}
    \mathcal{A}( \MsquaredKSpiRS, \MsquaredKSpiWS ) = \kern-2em \sum\limits_{r \neq (K\pi/\pi\pi)_{L=0} }^{} \kern-2em a_{r} e^{i \phi_{r}} \mathcal{A}_{r} ( \MsquaredKSpiRS, \MsquaredKSpiWS ) + F_{1} (\Msquaredpipi) + \mathcal{A}_{{K\pi}_{L=0}} (\MsquaredKSpiRS) + \mathcal{A}_{{K\pi}_{L=0}} (\MsquaredKSpiWS) \mathrm{.} \label{eqn:Dalitz_amplitude_model}  
\end{equation}
\end{widetext}
The $\pi\pi$ and $K\pi$ contributions with non-zero angular momentum are parameterized in the isobar ansatz by a coherent sum of the contributing intermediate quasi-two-body amplitudes.
In the coherent sum, the $r^{\mathrm{th}}$ intermediate quasi-two-body amplitude $\mathcal{A}_{r}$ enters with magnitude $a_{r}$ and relative phase $\phi_{r}$.
The symbol $F_{1}$ denotes the decay amplitude for the $\pi\pi$ $S$-wave contributions parameterized by the $K$-matrix approach,
and the symbol $\mathcal{A}_{{K\pi}_{L=0}}$ denotes the amplitude for the $K\pi$ $S$-wave contribution using the LASS parametrization.

\paragraph{Isobar ansatz.}
In the isobar ansatz, the quasi-two-body amplitude for a neutral \D meson decaying via the $r^{\mathrm{th}}$ intermediate resonance $(h_{1}h_{2})_{r}$ with spin $L$ to the three-body final state $h_{1}h_{2}h_{3}$ can be written as
\begin{align}
  A_{r}( \MsquaredKSpiRS, \MsquaredKSpiWS ) = & F_{D}^{(L)} (q, q_{0}) \times F_{r}^{(L)} (p, p_{0}) \nonumber \\ & \times Z_{L}(\Omega) \times T_{r} (m) \mathrm{,} \label{eqn:isobar_model_resonance_decay_amplitude}
\end{align}
where the terms are described below.

The form factors $F_{D}^{(L)}$ and $F_{r}^{(L)}$ describe the production $D \to r h_{3}$ and the decay $r \to h_{1}h_{2}$ of the resonance $r$ and the daughters of the resonance, respectively.
The form factors are parameterized by Blatt-Weisskopf barrier penetration factors~\cite{BlattWeisskopfFormFactors} that account for spin-dependent effects and prevent the decay amplitudes from diverging for large momentum transfers.
The factors depend on the momentum $q$ ($p$) of the bachelor particle $h_{3}$ (one of the resonance's daughter particles $h_{1}$ or $h_{2}$)
evaluated in the resonance rest frame, and $q_{0}$ ($p_{0}$) is the value of $q$ ($p$) when the invariant mass equals the pole mass of the resonance.
The Blatt-Weisskopf barrier penetration factors are defined as
\begin{align}
   L =\ & 0 : F^{(0)} (z, z_{0}) = 1 \mathrm{,} \label{eqn:Blatt_Weisskopf_barrier_factors_for_L_equal_zero} \\
   L =\ & 1 : F^{(1)} (z, z_{0}) = \sqrt{ \frac{ 1 + z_{0} }{ 1 + z } } \mathrm{,} \label{eqn:Blatt_Weisskopf_barrier_factors_for_L_equal_one} \\
   L =\ & 2 : F^{(2)} (z, z_{0}) = \sqrt{ \frac{ \left(z_{0} - 3\right)^{2} + 9 z_{0} }{ \left(z - 3\right)^{2} + 9 z } } \mathrm{,} \label{eqn:Blatt_Weisskopf_barrier_factors_for_L_equal_two}
\end{align}
where $z = \left( \lvert q \rvert d \right)^{2}$ and $z_{0} = \left( \lvert q_{0} \rvert d \right)^{2}$.
The parameter $d$ represents the meson radius or the impact parameter of the decay particles for the \D meson $d_{\D}$ and the resonances $d_{r}$, respectively.
In the present analysis, $d_{\D} = 5 \, \hbar c / \mathrm{GeV} \approx 1 \, \mathrm{fm}$ and $d_{r} = 1.5 \, \hbar c / \mathrm{GeV} \approx 0.3 \, \mathrm{fm}$ are applied.

The Zemach formalism~\cite{ZemachTensors} allows to describe the angular components of the amplitudes in a spin-tensor approach.
The Zemach tensor formalism is applied to express the angular correlations among the final state particles by the function $Z_{L}(\Omega)$, where the symbol $\Omega$ represent the angular relations of the involved particles.

The propagator term $T_{r}$ describes the dynamics in the resonance decay.
In the present analysis, the term is parameterized by a relativistic Breit-Wigner (BW) lineshape function defined as
\begin{equation}
  T_{r}( m ) = \frac{1}{ m_{0}^{2} - m^{2} - i m_{0} \Gamma \left(m\right) } \mathrm{,} \label{eqn:Breit_Wigner_lineshape}
\end{equation}
where $m_{0}$ denotes the pole mass of the resonance, and the mass-dependent width $\Gamma$ is given by
\begin{equation}
  \Gamma \left( m \right) = \Gamma_{0} \left( \frac{q}{q_{0}} \right)^{\left( 2L+1 \right)} \left( \frac{m_{0}}{m} \right) {F_{r}^{(L)}}^{2} \mathrm{.} \label{eqn:massdependent_width}
\end{equation}

The isobar ansatz is applied to parameterize the $P$- and $D$-wave contributions to the \DzerotoKSpipi decay.
In the nominal Dalitz plot amplitude model, the following intermediate quasi-two-body resonances are included:
the Cabibbo-favored
$K^{*}(892)^{-} \pip$,
$K^{*}_{2}(1430)^{-} \pip$,
$K^{*}(1680)^{-} \pip$,
$K^{*}(1410)^{-} \pip$ channels;
the doubly Cabibbo-suppressed
$K^{*}(892)^{+} \pim$,
$K^{*}_{2}(1430)^{+} \pim$,
$K^{*}(1410)^{+} \pim$ modes;
and the \CP eigenstates
$\KS \rho(770)^{0}$,
$\KS \omega(782)$,
$\KS f_{2}(1270)$,
and
$\KS \rho(1450)^{0}$.
To reduce the complexity of the Dalitz plot amplitude analysis, the masses and widths are fixed to the world averages~\cite{PDG2016} for all resonances except for the $K^{*}(892)^{\pm}$, whose values are measured in the fit.

\paragraph{$K$-matrix formalism.}
The isobar ansatz has limitations, for example, in the case of broad and overlapping resonances or for resonances located close to thresholds of additional decay channels~\cite{ReviewDalitzPlotAnalysis}.
An alternative approach is provided by the $K$-matrix formalism, which preserves unitarity by construction in the presence of overlapping resonances and coupled channels.
The $K$-matrix formalism is particularly suitable to describe the $J^{PC} = 0^{++}$ scalar contributions to the complex $S$-wave dynamics occurring in the $\pip\pim$ system of \DzerotoKSpipi decays.
The \babar, Belle, and LHCb experiments previously employed the $K$-matrix approach in Dalitz plot amplitude analyses of \DzerotoKSpipi decays to
perform measurements of \Dz-\Dzb oscillations~\cite{BABAR2010,Peng2014} and measurements of the Unitarity Triangle angle $\gamma$~\cite{phi_three} in \B meson decays~\cite{BABAR2008,LHCb2014}.
Following the previous measurements, the $K$-matrix formalism in the $P$-vector approximation~\cite{Aitchison1972} is applied to model the $\pi\pi$ $S$-wave contribution to the \DzerotoKSpipi decay.

In this parametrization, the decay amplitude $F_{1}$ entering in Eq.~(\ref{eqn:Dalitz_amplitude_model}) as the contribution of the $\pi\pi$ $S$-wave is defined by the relation
\begin{equation}
   F_{i} (s) = \left[ I - i K (s) \rho (s) \right]_{ij}^{-1} P_{j} (s) \mathrm{,} \label{eqn:Kmatrix_decay_amplitude}
\end{equation}
where the indices $i$ and $j$ denote the particular channels ($1 = \pi\pi$, $2 = K\bar{K}$, $3 = \pi\pi\pi\pi$, $4 = \eta\eta$, and $5 = \eta\eta'$) contributing to the scattering process.
The production vector $P$ parameterizes the initial production of states into the open channels, and the $K$-matrix describes the scattering process.
In this analysis, only the $\pip\pim$ final states are considered, and $s$ is the square of the invariant mass of the $\pip\pim$ system.
The terms $I$ and $\rho$ are the identity matrix and the phase-space matrix, respectively.
The $K$-matrix is defined as
\begin{equation}
   K_{ij} (s) = \left( f_{ij}^{\mathrm{scatt}} \frac{1 - s_{0}^{\mathrm{scatt}}}{s - s_{0}^{\mathrm{scatt}}} + \sum\limits_{\alpha}^{} \frac{g_{i}^{\alpha} g_{j}^{\alpha}}{m_{\alpha}^{2} - s} \right) f_{A0} (s) \mathrm{.} \label{eqn:Kmatrix_definition}
\end{equation}
The parameters $m_{\alpha}$ are the physical poles of the $K$-matrix, while $g_{i}^{\alpha}$ are the coupling constants of the $i$-th channel to the pole $\alpha$.
The parameters $f_{ij}^{\mathrm{scatt}}$ and $s_{0}^{\mathrm{scatt}}$ describe the smooth part of the $K$-matrix that is slowly varying.
The unit of the number 1 is in $\mathrm{GeV}/c^{2}$. The symbol $f_{A0}$ is the so-called ``Adler zero'' factor, defined as:
\begin{equation}
  f_{A0} (s) =  \frac{1 - s_{A0}}{s - s_{A0}} \left( s - s_{A} \frac{m_{\pi}^{2}}{2} \right) \mathrm{.}
\end{equation}
This factor suppresses the false kinematic singularity at $s = 0$ in the physical region close to the $\pip\pim$ threshold~\cite{Adler1965}.

The production vector $P$ has the same pole structure as the $K$-matrix and is defined as:
\begin{equation}
   P_{j} (s) = f_{1j}^{\mathrm{prod}} \frac{1 - s_{0}^{\mathrm{prod}}}{s - s_{0}^{\mathrm{prod}}} + \sum\limits_{\alpha}^{} \frac{\beta_{\alpha} g_{j}^{\alpha}}{m_{\alpha}^{2} - s} \mathrm{.}
\end{equation}
The $\beta_{\alpha}$ are the complex production couplings, and the parameters $f_{1j}^{\mathrm{prod}}$ and $s_{0}^{\mathrm{prod}}$
describe the production of the slowly-varying part of the $K$-matrix.

%%%%%%%%%%%%%%%%%%%%%%%%%%%%%%%%%%%%%%%%%%%%%%%%%%%%%%%%%%%%%%%%%%%%%%%%%%%%%%%%%%%%%%%%%%%%%%%%%%%%%%%%%%%%%%%%
%%% Table: K-Matrix Parameters
%%%%%%%%%%%%%%%%%%%%%%%%%%%%%%%%%%%%%%%%%%%%%%%%%%%%%%%%%%%%%%%%%%%%%%%%%%%%%%%%%%%%%%%%%%%%%%%%%%%%%%%%%%%%%%%%
\begin{table*}[htb]
\caption{The $K$-matrix parameters estimated by a global analysis of available $\pi\pi$ scattering data
(taken from Refs.~\cite{Kmatrix2003,BABAR2008}). The units of the pole masses $m_{\alpha}$ and the coupling
constants $g_{i}^{\alpha}$ are in \gevcc. The units of $s_{0}^{\mathrm{scatt}}$ and $s_{A0}$ are \gevgevcccc, while $s_{A}$ is dimensionless.}
\label{table:KMatrixParameters}
% \begin{tabular}{@{\hspace{0.5cm}}l@{\hspace{0.5cm}}  @{\hspace{0.5cm}}c@{\hspace{0.5cm}}  @{\hspace{0.5cm}}c@{\hspace{0.5cm}}  @{\hspace{0.5cm}}c@{\hspace{0.5cm}}  @{\hspace{0.5cm}}c@{\hspace{0.5cm}}  @{\hspace{0.5cm}}c@{\hspace{0.5cm}}}
\begin{tabular}{ D{,}{.}{-1} D{,}{.}{-1} D{,}{.}{-1} D{,}{.}{-1} D{,}{.}{-1} D{,}{.}{-1} }
\hline \hline
\multicolumn{1}{r}{$m_{\alpha}$} & \multicolumn{1}{r}{$g_{\pip\pim}^{\alpha}$} & \multicolumn{1}{r}{$g_{K\bar{K}}^{\alpha}$} & \multicolumn{1}{r}{$g_{4\pi}^{\alpha}$} & \multicolumn{1}{r}{$g_{\eta\eta}^{\alpha}$} & \multicolumn{1}{r}{$g_{\eta\eta^{\prime}}^{\alpha}$} \\
\hline
0,65100 & 0,22889 & -0,55377 &  0,00000 & -0,39899 & -0,34639 \\
1,20360 & 0,94128 &  0,55095 &  0,00000 &  0,39065 &  0,31503 \\
1,55817 & 0,36856 &  0,23888 &  0,55639 &  0,18340 &  0,18681 \\
1,21000 & 0,33650 &  0,40907 &  0,85679 &  0,19906 & -0,00984 \\
1,82206 & 0,18171 & -0,17558 & -0,79658 & -0,00355 &  0,22358 \\
\hline
& \multicolumn{1}{r}{$f_{11}^{\mathrm{scatt}}$} & \multicolumn{1}{r}{$f_{12}^{\mathrm{scatt}}$} & \multicolumn{1}{r}{$f_{13}^{\mathrm{scatt}}$} & \multicolumn{1}{r}{$f_{14}^{\mathrm{scatt}}$} & \multicolumn{1}{r}{$f_{15}^{\mathrm{scatt}}$} \\
& 0,23399 & 0,15044 & -0,20545 & 0,32825 & 0,35412 \\
\hline
\multicolumn{1}{r}{$s_{0}^{\mathrm{scatt}}$} & \multicolumn{1}{r}{$s_{A0}$} & \multicolumn{1}{r}{$s_{A}$} \\
-3,92637 & \multicolumn{1}{r}{$-0.15$} & \multicolumn{1}{r}{1} \\
\hline \hline
\end{tabular}
\end{table*}

In the present analysis, the $K$-matrix parameters $m_{\alpha}$, $g_{i}^{\alpha}$, $f_{ij}^{\mathrm{scatt}}$, $s_{0}^{\mathrm{scatt}}$, $s_{A0}$, and $s_{A}$ are fixed to the results
of a global analysis of available $\pi\pi$ scattering data~\cite{Kmatrix2003,BABAR2008} as summarized in Table~\ref{table:KMatrixParameters}.
The complex production couplings $\beta_{\alpha}$ and the production parameters $f_{1j}^{\mathrm{prod}}$ are free parameters determined from the fit.

\paragraph{LASS parametrization.}
For the $K\pi$ $S$-wave, an approach introduced by the LASS collaboration to describe $\Km\pip$ scattering processes is applied~\cite{LASS}.
The Cabibbo-favored $K^{*}_{0}(1430)^{-}$ and the doubly Cabibbo-suppressed $K^{*}_{0}(1430)^{+}$ contributions are each described by the empirical LASS parametrization.
The LASS parametrization is constructed from a BW term for the $K^{*}_{0}(1430)$ and a nonresonant component that has an effective range and introduces a phase shift:
\begin{equation}
  \mathcal{A}_{K\pi_{L=0}} (s) = R \sin \delta_{R} e^{i \delta_{R}} e^{i 2\delta_{F}} + F \sin \delta_{F} e^{i \delta_{F}} \mathrm{,} \label{eqn:LASS_contribution}
\end{equation}
where
\begin{align}
   \delta_{R} =& \phi_{R} + \tan^{-1} \left[ \frac{ M \Gamma( m_{K\pi}^{2} ) }{M^{2} - m_{K\pi}^{2} } \right] \mathrm{,} \label{eqn:LASS_delta_R} \\
   \delta_{F} =& \phi_{F} + \cot^{-1} \left[ \frac{1}{a q} + \frac{r q}{2} \right] \mathrm{.} \label{eqn:LASS_delta_F}
\end{align}
The parameters $R$ ($\phi_{R}$) and $F$ ($\phi_{F}$) are the amplitudes (phases) of the resonant and nonresonant components, respectively.
The parameters $a$ and $r$ are the scattering length and the effective interaction length, and $q$ represents the momentum of the spectator particle in the $K\pi$ rest frame.
The parameters $M$ and $\Gamma( M_{K\pi}^{2} )$ are the mass and the mass-dependent width of the resonant term defined in Eq.~(\ref{eqn:massdependent_width}),
and the phases $\delta_{R}$ and $\delta_{F}$ depend on $m_{K\pi}^{2}$.
According to Ref.~\cite{BABAR2008}, this parametrization is equivalent to a $K$-matrix approach
that describes a rapid phase shift originating from the resonant term and a slowly rising phase shift originating from the nonresonant term.
The mass and the width of the $K^{*}_{0}(1430)^{\pm}$ and the LASS $R$, $\phi_{R}$, $F$, $\phi_{F}$, $a$, and $r$ are free parameters measured in the fit.
The LASS parameters are required to be the same for the Cabibbo-favored $K^{*}_{0}(1430)^{-}$ and the doubly Cabibbo-suppressed $K^{*}_{0}(1430)^{+}$ contributions.

\subsubsection{Dalitz plot reconstruction efficiency correction}
\label{sec:D_decay_Dalitz_plot_reconstruction_efficiency_correction}
Experimental effects, for example from the detector acceptance, the reconstruction algorithms, or the event selection, can induce non-uniformities for the reconstruction efficiency as a function of the Dalitz plot phase space, $\epsilon ( \MsquaredKSpiRS, \MsquaredKSpiWS )$.
To account for these effects in the Dalitz amplitude analysis, the efficiency variations are estimated using a high-statistics sample of MC events of inclusive ${e^{+}e^{-} \to c\bar{c}}$ decays that contain
the \DstarplustoDzeropisoft, with \DzerotoKSpipi, signal decays.
In the MC simulations, the \DzerotoKSpipi decay is generated uniformly in the available \D meson decay phase space to uniformly populate the Dalitz plot.
The generated decays are passed to a GEANT3-based simulation with a specific Belle configuration to simulate the detector response.

The simulated detector response then undergoes the same reconstruction algorithms and event selection requirements as for the data.
The generated MC sample contains $50 \times 10^{6}$ \DstarplustoDzeropisoft, \DzerotoKSpipi signal decays, approximately 50 times the signal size in data, which enables the construction of a detailed map of the reconstruction efficiency.

The efficiency map is constructed using an approach \babar\ introduced in the search for the $Z(4430)^{-}$ state~\cite{BABAR2009_Dalitz_efficiency}.
In this approach, the efficiency is expressed as a function of the square of the two-body invariant mass \MsquaredKSpiRS and $\cos{ \theta_{\KS} }$.
The variable $\cos{ \theta_{\KS} }$ is computed by the normalized dot product between the $\KS \pi^{-}$ three-momentum vector measured in the \D meson rest frame and the three-momentum vector of the \KS meson
after a Lorentz transformation from the \D meson rest frame to the $\KS \pi^{-}$ rest frame.
This choice of variables naturally introduces a ``rectangular Dalitz plot'' that is insensitive to potential binning effects that may arise at the curved edges of the \MsquaredKSpiRS and \Msquaredpipi Dalitz phase space due to the finite MC sample statistics.
In order to parameterize the reconstruction efficiency and to smooth statistical fluctuations, the efficiency map is constructed in the following way.

In the first step, the angular variations of the efficiency are estimated by expanding the $\cos{ \theta_{\KS} }$ distributions by a linear combination of Legendre polynomials up to order $L=7$:
\begin{equation}
\epsilon ( \cos{ \theta_{\KS} } ) = \sum\limits_{L=0}^{7} c_{L} ( \MsquaredKSpiRS ) Y_{L}^{0} ( \cos{ \theta_{\KS} } ) \mathrm{.} \label{equation:efficiency_legendre_polynomial_expansion}
\end{equation}
The mass-squared dependent coefficients $c_{L}$ are estimated by fitting the linear combination of Legendre polynomials to the $\cos{ \theta_{\KS} }$ distributions in intervals of \MsquaredKSpiRS.
For each of the eight coefficients $c_{0}$, $c_{1}$, ..., $c_{7}$, this forms a distribution as a function of \MsquaredKSpiRS.
In the second step, each of the $c_{L}$ distributions is fitted as a function of \MsquaredKSpiRS.
The coefficient $c_{0}$ is modeled by a $5^{\mathrm{th}}$-order polynomial function multiplied with a sigmoid function. This choice of parametrization
enables us to properly describe the drop in the reconstruction efficiency near the upper boundary of \MsquaredKSpiRS.
The coefficients $c_{1}$, $c_{2}$, ..., $c_{7}$ are fitted by $5^{\mathrm{th}}$-order Chebyshev polynomial functions.

The chosen order for the polynomial functions has been found to be sufficient to describe the details of the efficiency variations and at the same time to be low enough to avoid overfitting any structures.
The dependence on the chosen order of the expansion in linear combinations of Legendre polynomials is weak, and lower or higher choices than $L=7$ yield consistent results.

\begin{figure*}
\includegraphics[width=0.85\textwidth]{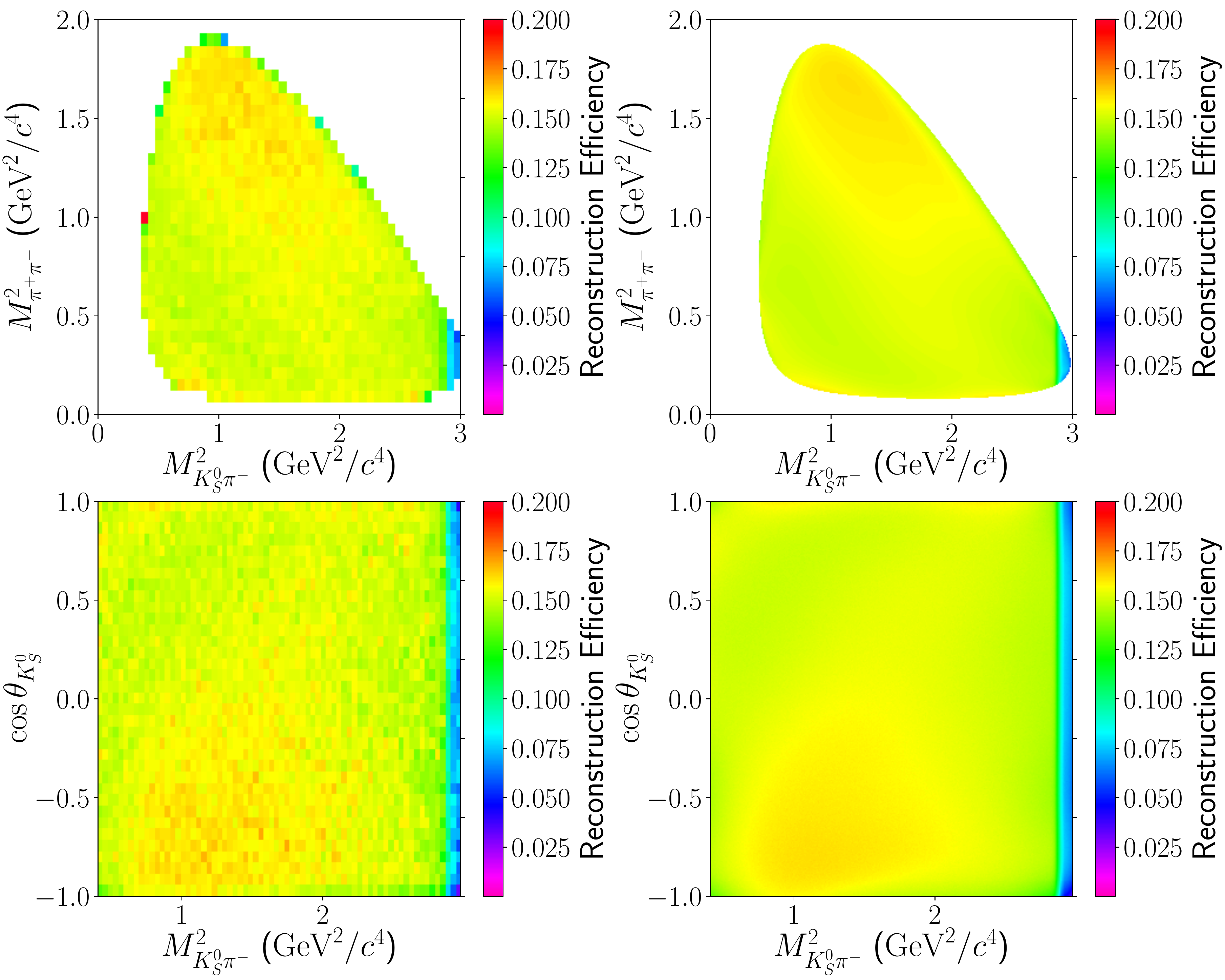}
\caption{
Variation of the Dalitz plot reconstruction efficiency as a function of \MsquaredKSpiRS and \Msquaredpipi (top),
and as a function of \MsquaredKSpiRS and $\cos{ \theta_{\KS} }$ (bottom).
The efficiency variations are estimated using a high-statistics sample of Monte Carlo events of inclusive ${e^{+}e^{-} \to c\bar{c}}$ decays
containing \DstarplustoDzeropisoft with \DzerotoKSpipi signal decays (left), and detailed efficiency maps (right) are constructed
by the parameterized model described in Sect.~\ref{sec:D_decay_Dalitz_plot_reconstruction_efficiency_correction}.
}
\label{fig:BELLE_Dalitz_plot_reconstruction_efficiency_parameterization_using_Legendre_model}
\end{figure*}

The reconstruction efficiency is almost flat over large parts of the Dalitz plot phase space. The efficiency decreases slightly at larger values of \MsquaredKSpiRS and drops close to the kinematic border.
The two-dimensional binned distributions of the reconstruction efficiency and the obtained parameterized efficiency maps are shown as a function of \MsquaredKSpiRS and \Msquaredpipi, and of \MsquaredKSpiRS and $\cos{ \theta_{\KS} }$,
in Fig.~\ref{fig:BELLE_Dalitz_plot_reconstruction_efficiency_parameterization_using_Legendre_model}.
The efficiency map represents the variations of the reconstruction efficiency well over the full Dalitz plot phase space, including the efficiency drops at the kinematic edges of the Dalitz plot.
The binned distributions of the reconstruction efficiency are compared to the parameterized efficiency map, and a reduced $\chi^{2}$ of $1.03$ is obtained for $2450$ degrees of freedom (d.o.f.).

%%%%%%%%%%%%%%%%%%%%%%%%%%%%%%%%%%%%%%%%%%%%%%%%%%%%%%%%%%%%%%%%%%%%%%%%%%%%%%%%%%%%%%%%%%%%%%%%%%%%%%%%%
%%% Figure: Dalitz Plot Reconstruction Efficiency
%%%%%%%%%%%%%%%%%%%%%%%%%%%%%%%%%%%%%%%%%%%%%%%%%%%%%%%%%%%%%%%%%%%%%%%%%%%%%%%%%%%%%%%%%%%%%%%%%%%%%%%%%

\subsubsection{Dalitz plot background description}
\label{sec:D_decay_Dalitz_plot_background_description}
The Dalitz plot distributions of the background are estimated from the data using two \MDzero--\deltaMDstarDzero sideband regions defined by
$1.815 < \MDzero < 1.835\,\gevcc$ and $150.4 < \deltaMDstarDzero < 160\,\mevcc$, and $1.895 < \MDzero < 1.915\,\gevcc$ and $150.4 < \deltaMDstarDzero < 160\,\mevcc$.
The distribution of the background has a smooth shape over the Dalitz plot. The background exhibits small resonant contributions from the $K^*(892)^-$, $K^*(1680)^-$, and $\rho(770)$ resonances,
and further contributions from the $K_0^*(1430)^-$, $K_2^*(1430)^-$, and $K^*(1410)^-$ resonances, which appear as a single broad enhancement.
In order to reduce the sensitivity to statistical fluctuations due to the finite sample statistics in the data sideband regions, a parameterized model of the background is constructed and fitted to the Dalitz plot distributions in the sidebands.
The background model is composed of a $6^{\mathrm{th}}$-order polynomial function for the smooth distributions
and BW lineshapes for the $K^*(892)^-$, $K^*(1680)^-$, and $\rho(770)^{0}$ resonances and for the mixture of excited kaon states at approximately $1410\,\mevcc$.
These resonant contributions are added incoherently.
The background model provides an accurate description of the background in all regions of the Dalitz plot phase space.\\

\subsubsection{Likelihood function and procedure for the $\DzerotoKSpipi$ Dalitz plot fit}
\label{sec:D_decay_Dalitz_plot_amplitude_analysis}
The \DzerotoKSpipi decay amplitude parameters are estimated by an unbinned ML fit to the Dalitz plot distributions of the flavor-tagged \Dz sample.
The likelihood function accounting for the contributions of the signal and background is written as
\begin{widetext}
\begin{align}
  \mathcal{L} = \prod\limits_{i=1}^{N} & \left[ f_{\mathrm{sig}} \times p_{\mathrm{sig}} ( \MsquaredKSpiRS, \MsquaredKSpiWS ) \right. \nonumber \\
  & \left. + \left(1 - f_{\mathrm{sig}} \right) \times \left( f_{\mathrm{rnd}} \times p_{\mathrm{rnd}} ( \MsquaredKSpiRS, \MsquaredKSpiWS ) + \left(1 - f_{\mathrm{rnd}} \right) \times p_{\mathrm{bkg}} ( \MsquaredKSpiRS, \MsquaredKSpiWS ) \right) \right] \mathrm{,} \label{equation:Dalitz_fit_definition_likelihood_function}
\end{align}
\end{widetext}
where the index $i$ runs over the reconstructed \DzerotoKSpipi candidates.
The signal fraction $f_{\mathrm{sig}}$ and the fraction of the random slow pion background $f_{\mathrm{rnd}}$ are determined by the two-dimensional fit to the \MDzero and \deltaMDstarDzero distributions.
The functions $p_{\mathrm{sig}}$, $p_{\mathrm{rnd}}$, and $p_{\mathrm{bkg}}$ are the p.d.f.s of the Dalitz plot distributions for the signal, the random slow pion background, and the remaining background, respectively.
The signal p.d.f.\  is constructed from the efficiency-corrected Dalitz plot intensities, computed from the absolute square of the \DzerotoKSpipi decay amplitude $\mathcal{A}( \MsquaredKSpiRS, \MsquaredKSpiWS )$ defined in Eq.~(\ref{eqn:Dalitz_amplitude_model}),
and by normalizing to the available Dalitz plot phase space:
\begin{widetext}
\begin{equation}
  p_{\mathrm{sig}} ( \MsquaredKSpiRS, \MsquaredKSpiWS ) = \frac{ \epsilon ( \MsquaredKSpiRS, \MsquaredKSpiWS ) \Bigl\lvert \mathcal{A}( \MsquaredKSpiRS, \MsquaredKSpiWS ) \Bigr\rvert^{2} }{ \mathlarger{\int}\limits_{D}^{} \epsilon ( \MsquaredKSpiRS, \MsquaredKSpiWS ) \Bigl\lvert \mathcal{A}( \MsquaredKSpiRS, \MsquaredKSpiWS ) \Bigr\rvert^{2} d \MsquaredKSpiRS d \MsquaredKSpiWS } \mathrm{.} \label{eqn:Dalitz_fit_pdf_signal}
\end{equation}
\end{widetext}
The random slow pion background is composed of a mixture of real \Dz and \Dzb mesons decaying to the $\KS\pip\pim$ final state.
During the reconstruction of \DstarplustoDzeropisoft decays, these \D mesons are combined with random slow pion candidates.
If the slow pion has the incorrect charge, the $c$-flavor content of the neutral \D meson will be misidentified and the wrong flavor will be assigned.
Neglecting possible production or detection asymmetries, the na\"{\i}ve expectation of the probability to select a slow pion track with the wrong charge is $p = 0.5$.
The decay amplitudes for \Dz and \Dzb mesons are related by an exchange of the Dalitz plot variables, $\mathcal{A}_{\Dz} = \mathcal{A}( \MsquaredKSpiRS, \MsquaredKSpiWS ) \leftrightarrow \mathcal{A}_{\Dzb} = \mathcal{A}( \MsquaredKSpiWS, \MsquaredKSpiRS )$.
The p.d.f.\  of the random slow pion background is constructed from the signal p.d.f.\  by allowing for the exchange of the Dalitz plot positions and is defined as
\begin{widetext}
\begin{equation}
   p_{\mathrm{rnd}} ( \MsquaredKSpiRS, \MsquaredKSpiWS ) = \left( 1 - f_{\mathrm{wtag}} \right) \times p_{\mathrm{sig}} ( \MsquaredKSpiRS, \MsquaredKSpiWS ) + f_{\mathrm{wtag}} \times p_{\mathrm{sig}} ( \MsquaredKSpiWS, \MsquaredKSpiRS ) \mathrm{.} \label{eqn:random_pion_background}
\end{equation}
\end{widetext}

The $f_{\mathrm{wtag}}$ quantifies the fraction of ``wrong \D meson flavor-tags'' and is estimated directly from the data by a separate Dalitz plot fit to the $150 < \deltaMDstarDzero < 155\,\mevcc$ sideband region
that has an enhanced population from the random slow pion background and no signal. In this Dalitz plot fit to the data sideband, the fraction of wrong \D meson flavor-tag is measured and the result is $f_{\mathrm{wtag}} = 0.492 \pm 0.075$,
in agreement with the na\"{\i}ve expectation. In the subsequent Dalitz plot fit to the signal region, $f_{\mathrm{wtag}}$ is fixed to the estimate obtained from the sideband.

The background p.d.f.\  $p_{\mathrm{bkg}}$ is constructed from the parameterized background model described in Sect.~\ref{sec:D_decay_Dalitz_plot_background_description}.
The background is composed of combinatorial background and additional contributions from processes containing real slow pions and wrong \Dz mesons.

Due to the high statistics of the Belle ${e^{+}e^{-} \to c\bar{c}}$ data sample of more than $10^{6}$ events, and the complexity of the \DzerotoKSpipi decay amplitude model,
maximizing the likelihood function and performing the Dalitz plot fit is computationally intensive, taking hours to days on a single CPU core of a recent Intel Xeon processor-based Linux workstation.
A new software framework for Dalitz plot amplitude analyses has been developed to increase the performance of the fit and to realize the present analysis.
Key features of the framework are the parallel computing algorithms for both the evaluation of the likelihood function defined in Eq.~(\ref{equation:Dalitz_fit_definition_likelihood_function}), and for the
numeric integration of the p.d.f.s.
The parallel computing algorithms are realized using OpenMP~\cite{dagum1998openmp,OpenMP} and enable the Dalitz plot fits to make simultaneous use of multiple CPUs to significantly reduce the required run time.
In the present analysis, a speed-up of approximately a factor of $40$ has been achieved for the time needed to reach convergence of the fit by using 64 CPU cores.

The \DzerotoKSpipi decay amplitude parameters are determined by maximizing Eq.~(\ref{equation:Dalitz_fit_definition_likelihood_function}) for the Dalitz plot distributions in the signal-enhanced region
defined in Sect.~\ref{sec:reconstruction_and_selection}.
The amplitude magnitudes $a_{r}$ and phases $\phi_{r}$ of the intermediate resonant states are free parameters in the fit,
and measured relative to the $\KS \rho(770)^{0}$ amplitude.
The $\KS \rho(770)^{0}$ amplitude is fixed to $a_{\KS \rho(770)^{0}}=1$ and $\phi_{\KS \rho(770)^{0}}=0^\circ$ and serves as a reference.

\subsubsection{Results of the $\DzerotoKSpipi$ Dalitz plot amplitude analysis}
\label{sec:D_decay_Dalitz_fit_results}
\begin{figure*}
\includegraphics[width=0.9\textwidth]{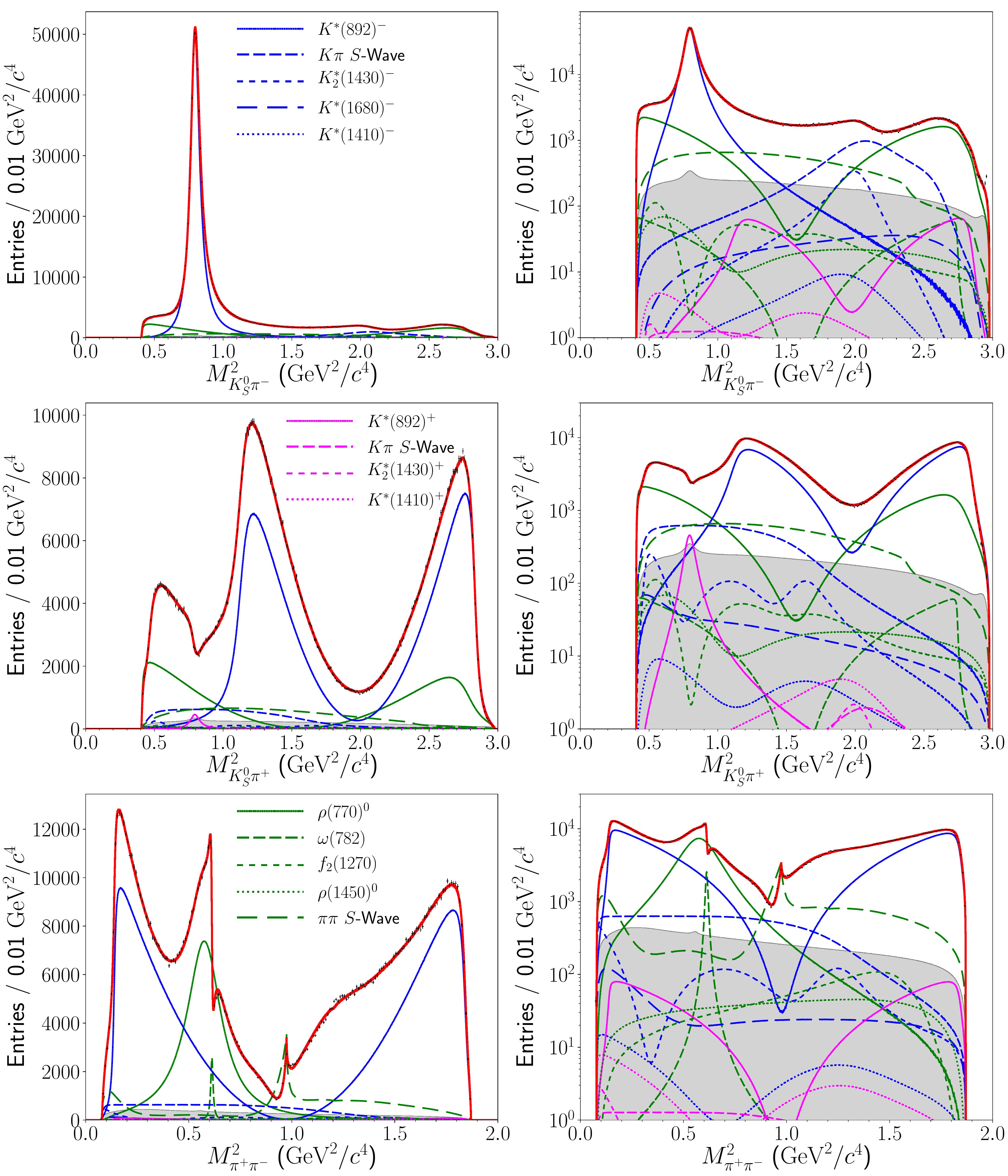}
\caption{
(color online). Projections of the Dalitz plot data distributions (points with error bars) for \DzerotoKSpipi from \DstarplustoDzeropisoft decays reconstructed from Belle ${e^{+}e^{-} \to c\bar{c}}$ data,
and of the result of the fit (lines). The red solid lines show the projections of the total fit function including background.
The dotted and dashed colored lines show projections of the individual components of the \DzerotoKSpipi decay amplitude model.
The blue, magenta, and green lines represent resonant and nonresonant contributions originating from the \MsquaredKSpiRS, \MsquaredKSpiWS, and \Msquaredpipi systems, respectively.
The left plots use a linear scale on the y-axis. The right plots show the same data distributions and fit projections with a log-scale in order to increase the visibility of components with very low fit fractions,
and other details of the model.
The components are computed from the squared amplitude of each intermediate resonant and nonresonant contribution scaled by its fit fraction. 
Various beautiful quantum mechanical phenomena can be observed: for example, the complex constructive and destructive interference patterns, and the dynamic generation
of the peak by the $K$-matrix formalism located close to the $f_{0}(980)$ in the \Msquaredpipi spectrum.
}
\label{figure_Dalitz_fit_components}
\end{figure*}

The results for the estimated \DzerotoKSpipi decay amplitude model parameters are summarized in Table~\ref{Dalitz_amplitude_model}.
The data distributions are shown in Figs.~\ref{figure_Dalitz_data_distributions} and~\ref{figure_Dalitz_fit_components}, and projections of the fit are shown in Fig.~\ref{figure_Dalitz_fit_components}.
The fit reproduces the data distributions well over the full range of the Dalitz plot.
The fit projections exhibit few deviations, for example, for the $\rho(770)^{0}$--$\omega(782)$ interference region in the $M^{2}_{\pi^{+}\pi^{-}}$ projection.
These deviations are very small compared to the overall scale of agreement.

The quality of the fit is estimated by a two-dimensional $\chi^{2}$ test. The Dalitz plot data distributions are binned into square intervals with an edge length of $0.01\,\mathrm{GeV}/c^{2}$ and then compared to the fit function.
A reduced $\chi^{2}$ of $1.05$ is obtained for $31\,272$ d.o.f. based on statistical uncertainties only,
indicating a good quality of the fit compared to previous models of this decay~\cite{BABAR2010,Peng2014,BABAR2008,Bellesigmatwo,CDFsigmatwo}.
The normalized residuals contributing to the $\chi^{2}$ function vary approximately uniformly over the Dalitz plot phase space and do not exhibit any macroscopic deviations or structures.

To quantify the contributions of individual amplitudes, the fit fractions ($FF$s) are evaluated. The $FF$ for the $r^{\mathrm{th}}$ intermediate resonant or nonresonant contribution is defined as:
\begin{equation}
  FF_{r} = \frac{ a_{r}^{2} \mathlarger{\int_{D}^{}} \Bigl\lvert \mathcal{A}_{r} ( \MsquaredKSpiRS, \MsquaredKSpiWS ) \Bigr\rvert^{2} d \MsquaredKSpiRS d \MsquaredKSpiWS }{ \mathlarger{\int_{D}^{}} \Bigl\lvert \mathcal{A}( \MsquaredKSpiRS, \MsquaredKSpiWS ) \Bigr\rvert^{2} d \MsquaredKSpiRS d \MsquaredKSpiWS } \mathrm{.} \label{eqn:fit_fraction}
\end{equation}
The sum of the fit fractions does not necessarily equal unity due to possible constructive or destructive interference effects among the amplitudes.
In the present Dalitz plot amplitude analysis, the total fit fraction is $101.6\%$.
The \DzerotoKSpipi decay is dominated by the $\Dz \to K^{*}(892)^{-}\pi^{+}$ mode which has a fit fraction of $59.9\%$.
The second largest contribution is $\Dz \to \KS \rho(770)^{0}$ with a fit fraction of $20.4\%$, followed by the $\pip\pim$ $S$-wave with $10.0\%$.

\begin{figure*}[htb]
\includegraphics[width=0.9\textwidth]{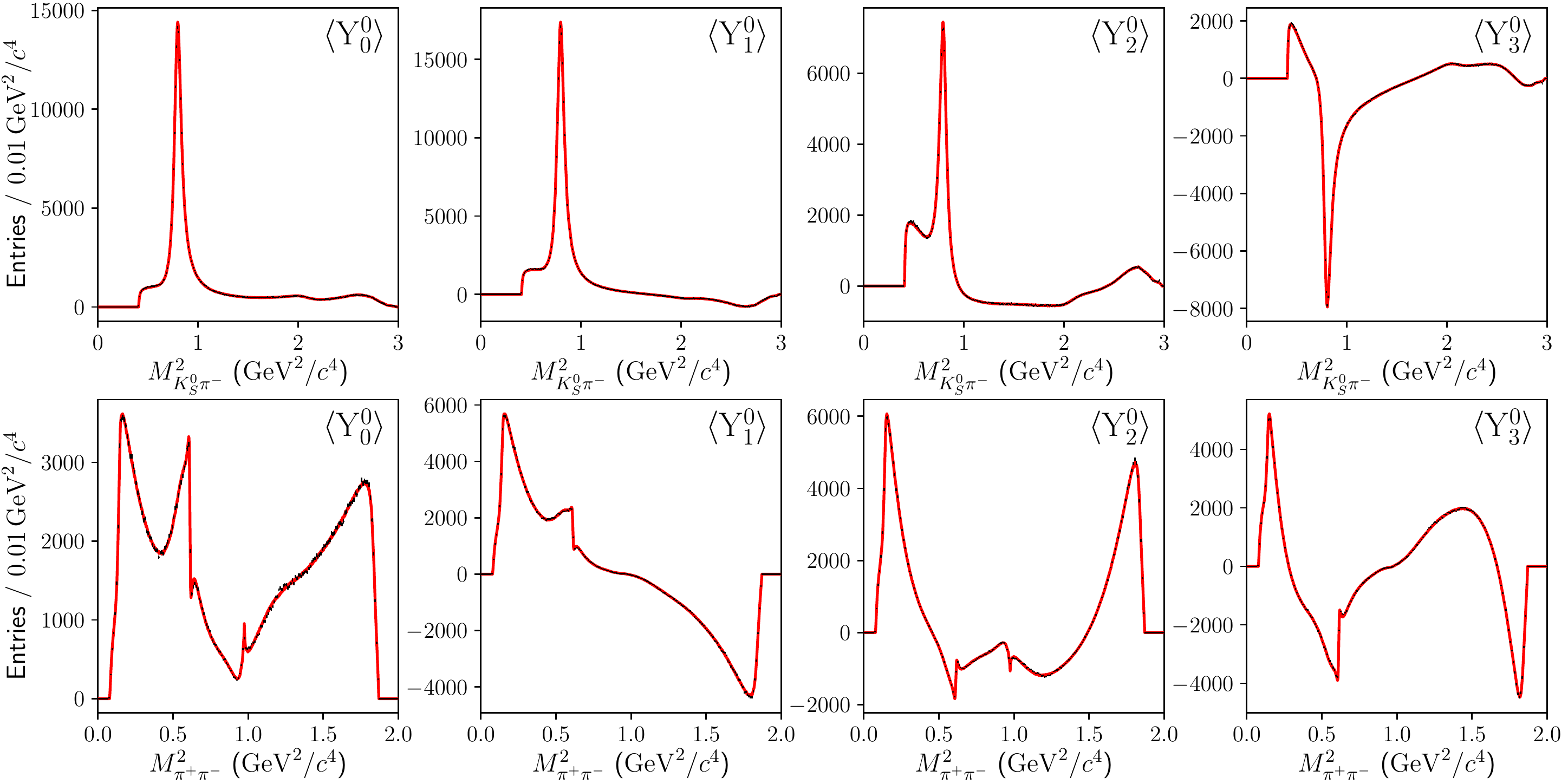}
\caption{
(color online). Dalitz plot data distributions (points with error bars) for \DzerotoKSpipi from \DstarplustoDzeropisoft decays reconstructed from Belle ${e^{+}e^{-} \to c\bar{c}}$ data,
and projections of the Dalitz plot fit (red solid lines) for \MsquaredKSpiRS (top) and \Msquaredpipi (bottom)
weighted by the corresponding Legendre moments.
}
\label{figure_Legendre_moments}
\end{figure*}

To test further the agreement of the Dalitz plot amplitude model with the data, we follow an approach employed by \babar\ in Ref.~\cite{BaBar2011}.
The Dalitz plot data distributions along the mass-squared directions are weighted by $Y_{k}^{0} (\cos \theta) = \sqrt{(2 k + 1) / 4 \pi} \, P_{k} (\cos \theta)$, where $P_{k}$ is the Legendre polynomial function of $k^{\mathrm{th}}$-order,
and compared to the expectation of the corresponding Legendre moment computed from the Dalitz plot amplitude model.
For \MsquaredKSpiRS  and \Msquaredpipi, the weighted data distributions and the Legendre moments up to the $3^{\mathrm{rd}}$-order are shown in Fig.~\ref{figure_Legendre_moments}.
The chosen representation is sensitive to the local phase and interference structures of the contributing amplitudes, complementary to the mass-squared projections.
Good agreement is observed between the data distributions and the Dalitz plot amplitude model.

%%%%%%%%%%%%%%%%%%%%%%%%%%%%%%%%%%%%%%%%%%%%%%%%%%%%%%%%%%%%%%%%%%%%%%%%%%%%%%%%%%%%%%%%%%%%%%%%%%%%%%%%%
%%% Figure: Dalitz fit components
%%%%%%%%%%%%%%%%%%%%%%%%%%%%%%%%%%%%%%%%%%%%%%%%%%%%%%%%%%%%%%%%%%%%%%%%%%%%%%%%%%%%%%%%%%%%%%%%%%%%%%%%%

%%%%%%%%%%%%%%%%%%%%%%%%%%%%%%%%%%%%%%%%%%%%%%%%%%%%%%%%%%%%%%%%%%%%%%%%%%%%%%%%%%%%%%%%%%%%%%%%%%%%%%%%%%%%%%%%
%%% Table: Results of Dalitz fit
%%%%%%%%%%%%%%%%%%%%%%%%%%%%%%%%%%%%%%%%%%%%%%%%%%%%%%%%%%%%%%%%%%%%%%%%%%%%%%%%%%%%%%%%%%%%%%%%%%%%%%%%%%%%%%%%
\begin{table*}
\caption{Results for the amplitude magnitudes $a_{r}$, phases $\phi_{r}$, fit fractions, $K$-matrix parameters for the $\pip\pim$ $S$-wave, LASS parameters for the $K\pi$ $S$-wave, and $K^{*}(892)^{\pm}$ parameters
determined by the \DzerotoKSpipi Dalitz plot fit performed for \DstarplustoDzeropisoft events reconstructed from Belle data.
Uncertainties are statistical only.}
\label{Dalitz_amplitude_model}
\begin{tabular}
 {@{\hspace{0.5cm}}l@{\hspace{0.5cm}}  @{\hspace{0.5cm}}r@{\hspace{0.5cm}}  @{\hspace{0.5cm}}r@{\hspace{0.5cm}}  @{\hspace{0.5cm}}r@{\hspace{0.5cm}}}
\hline \hline
	                \multicolumn{1}{l}{Resonance} & \multicolumn{1}{c}{Amplitude\phantom{0000000}} & \multicolumn{1}{c}{Phase (deg)\phantom{0000}} & \multicolumn{1}{c}{Fit Fraction (\%)} \\
\hline
$\KS \rho(770)^{0}$          &  $ 1\,\mathrm{(fixed)}$& $  0\,\mathrm{(fixed)} $               & $ 20.4 $ \\
$\KS \omega(782)$        &  $ 0.0388 \pm 0.0005 $ & $  120.7 \pm    0.7 $ & $ 0.5  $ \\
$\KS f_{2}(1270)$        &  $ 1.43 \pm 0.03\phantom{00} $     & $  -36.3 \pm    1.1 $ & $ 0.8  $ \\
$\KS \rho(1450)^{0}$         &  $ 2.85 \pm 0.10\phantom{00} $     & $  102.1 \pm    1.9 $ & $ 0.6  $ \\

$K^{*}(892)^{-} \pip$     &  $ 1.720 \pm 0.006\phantom{0} $   & $  136.8 \pm    0.2 $ & $ 59.9 $ \\
$K^{*}_{2}(1430)^{-} \pip$ &  $ 1.27 \pm 0.02\phantom{00} $     & $  -44.1 \pm    0.8 $ & $ 1.3  $ \\
$K^{*}(1680)^{-} \pip$    &  $ 3.31 \pm 0.20\phantom{00} $     & $ -118.2 \pm    3.1 $ & $ 0.5  $ \\
$K^{*}(1410)^{-} \pip$    &  $ 0.29 \pm 0.03\phantom{00} $     & $  99.4 \pm    5.5 $ & $ 0.1  $ \\

$K^{*}(892)^{+} \pim$     &  $ 0.164 \pm 0.003\phantom{0} $   & $  -42.2 \pm    0.9 $ & $ 0.6  $ \\
$K^{*}_{2}(1430)^{+} \pim$ &  $ 0.10 \pm 0.01\phantom{00} $     & $  -89.6 \pm    7.6 $ & $ <0.1 $ \\
$K^{*}(1410)^{+} \pim$    &  $ 0.21 \pm 0.02\phantom{00} $     & $  150.2 \pm    5.3 $ & $ <0.1 $ \\
\hline
\multicolumn{1}{l}{$\pip\pim$ $S$-wave Parameters} & & & $ 10.0 $ \\
$ \beta_{1} $          &      $   8.5 \pm  0.5\phantom{000} $  & $    68.5 \pm    3.4 $ & \\
$ \beta_{2} $          &      $  12.2 \pm  0.3\phantom{000} $  & $    24.0 \pm    1.4 $ & \\
$ \beta_{3} $          &      $  29.2 \pm  1.6\phantom{000} $  & $    -0.1 \pm    2.5 $ & \\
$ \beta_{4} $          &      $  10.8 \pm  0.5\phantom{000} $  & $   -51.9 \pm    2.4 $ & \\

$ f^{\mathrm{prod}}_{11} $      &   $   8.0 \pm  0.4\phantom{000} $  & $  -126.0 \pm    2.5 $ & \\
$ f^{\mathrm{prod}}_{12} $      &   $  26.3 \pm  1.6\phantom{000} $  & $  -152.3 \pm    3.0 $ & \\
$ f^{\mathrm{prod}}_{13} $      &   $  33.0 \pm  1.8\phantom{000} $  & $   -93.2 \pm    3.1 $ & \\
$ f^{\mathrm{prod}}_{14} $      &   $  26.2 \pm  1.3\phantom{000} $  & $  -121.4 \pm    2.7 $ & \\

$s^{\mathrm{prod}}_{0} $        &  $-0.07\,\mathrm{(fixed)}$ & & \\
\hline
\multicolumn{1}{l}{$K\pi$ $S$-wave Parameters} & & & \\
$K^{*}_{0}(1430)^{-} \pip$ &  $ 2.36 \pm 0.06\phantom{00} $     & $   99.4 \pm    1.7 $ & $ 7.0  $ \\
$K^{*}_{0}(1430)^{+} \pim$ &  $ 0.11 \pm 0.01\phantom{00} $     & $  162.3 \pm    6.6 $ & $ <0.1 $ \\
$ \mathrm{M}_{K^{*}_{0}(1430)^{\pm}} \,(\mathrm{GeV}/c^{2})$ & $ 1.441 \pm 0.002\phantom{0} $ & & \\
$ \Gamma_{K^{*}_{0}(1430)^{\pm}} \,(\mathrm{GeV})$     & $ 0.193 \pm 0.004\phantom{0} $ & & \\
$F$                         & $ +0.96 \pm    0.07\phantom{00} $ & & \\
$R$                         & $ 1\,\mathrm{(fixed)}$ & & \\
$a$                         & $ +0.113 \pm    0.006\phantom{0} $ & & \\
$r$                         & $ -33.8 \pm    1.8\phantom{000} $ & & \\
$\phi_{F} \,\mathrm{(deg)}$ & $   0.1 \pm    0.3\phantom{000} $ & & \\
$\phi_{R} \,\mathrm{(deg)}$ & $ -109.7 \pm   2.6\phantom{000} $ & & \\
\hline
\multicolumn{1}{l}{$ K^{*}(892)^{\pm}$ Parameters} & & & \\
$ \mathrm{M}_{K^{*}(892)^{\pm}} \,(\mathrm{GeV}/c^{2})$ & $ 0.8937 \pm 0.0001 $ & & \\
$ \Gamma_{K^{*}(892)^{\pm}} \,(\mathrm{GeV})$     & $ 0.0472 \pm 0.0001 $ & & \\
\hline \hline
\end{tabular}
\end{table*}

%%%%%%%%%%%%%%%%%%%%%%%%%%%%%%%%%%%%%%%%%%%%%%%%%%%%%%%%%%%%%%%%%%%%%%%%%%%%%%%%%%%%%%%%%%%%%%%%%%%%%%%%%
%%% Figure: Legendre moment weighted distributions
%%%%%%%%%%%%%%%%%%%%%%%%%%%%%%%%%%%%%%%%%%%%%%%%%%%%%%%%%%%%%%%%%%%%%%%%%%%%%%%%%%%%%%%%%%%%%%%%%%%%%%%%%

\subsubsection{Model variations and crosschecks}
\label{sec:D_decay_Dalitz_fit_validation}
The Dalitz plot amplitude analysis of \DzerotoKSpipi decays is validated by various crosschecks. Before choosing the nominal Dalitz plot amplitude model,
various alternative parameterizations and model variations have been considered.

The addition of further resonances (for example, the $K^{*}(1680)^{+} \pim$ mode) does not improve the fit quality nor result in significant fit fractions for these resonances.
When parameterizing the $\rho(770)^{0}$ resonance by the Gounaris-Sakurai lineshape function~\cite{GounarisSakurai} instead of the BW lineshape, worse agreement with the data is observed
for the $\rho(770)^{0}$ and the $\rho(770)^{0}$--$\omega(782)$ interference region.
The determination of more parameters in the Dalitz plot fit (for example, the mass and the width of the $\rho(770)^{0}$, $\omega(782)$, or other resonances) does not significantly improve the fit quality.
In the nominal model, these parameters are fixed to the world averages~\cite{PDG2016} in order to reduce the complexity of the Dalitz plot fit.

Instead of the $K$-matrix and the LASS parametrization to describe the $\pip\pim$ and $K\pi$ $S$-waves, a model based on a pure isobar approach has been considered.
In the isobar model, the $\pip\pim$ $S$-wave is modeled by the $\sigma_{1}$, $\sigma_{2}$, $f_{0}(980)$, and $f_{0}(1370)$ resonances,
and the $K\pi$ $S$-waves by the Cabibbo-favored $K^{*}_{0}(1430)^{-}$  and the doubly Cabibbo-suppressed $K^{*}(1410)^{+}$ resonances parameterized by BW lineshapes.
An additional term that is constant in phase space is added to account for nonresonant contributions.
For the isobar model, a reduced $\chi^{2}$ of $1.23$ is obtained for $31287$ d.o.f.
A similar isobar model including the $\sigma_{2}$ resonance has been used before by Belle~\cite{Bellesigmatwo2004,Bellesigmatwo} and CDF~\cite{CDFsigmatwo} in Dalitz plot amplitude analyses of \DzerotoKSpipi decays.
However, since the physical nature is not firmly established for all these states, in particular for the $\sigma_{2}$ resonance, and less agreement with the data was observed for the isobar model, it is not chosen as the nominal model.

The CLEO experiment performed a model-independent determination of the relative strong phase between $\Dz$ and $\Dzb \to \KS \pip \pim$ decays by exploiting the quantum correlation of $\Dz\Dzb$ pairs produced from $\psi(3770)$ decays
in $e^{+}e^{-}$ annihilations~\cite{CLEOstrongphases}. The results obtained in 8 bins of the Dalitz phase space are compared to the relative strong phase evaluated from the nominal Dalitz plot amplitude model.
Very good agreement with the model-independent measurement is observed, corresponding to a $p$-value of $0.46$.
The results also agree well with a previous \babar\ model of the same decay~\cite{BABAR2008} that has been applied by CLEO to optimize the binning for the model-independent measurement of the relative strong phase.

\section{Time-dependent Dalitz plot analysis of {\boldmath$\Bz \to \D^{(*)} \MakeLowercase{h^{0}}$} with {\boldmath$\D \to \KS \pip \pim$} decays using \babar\ and Belle data}
\label{sec:Bdecay_timedependent_Dalitz_plot_analysis}

\subsection{Event reconstruction and selection}
\label{sec:Bdecay_reconstruction_and_selection}
The similar performance of the \babar\ and Belle  detectors allows the use of almost identical selection requirements in the two data sets.
The event reconstruction and applied selection requirements discussed below follow the strategy used for the previous combined \babar+Belle analysis of $\Bz \to \D^{(*)}_{\CP} h^{0}$ decays described in Ref.~\cite{Roehrken2015}.

Charged pion candidates are formed from tracks that are reconstructed from detected hits inside the tracking detectors and meet criteria for charged particles~\cite{BaBarDetector,BelleDetector}.
Photons are reconstructed from energy deposits of electromagnetic showers detected in the electromagnetic calorimeters. The energy of a photon candidate is required to be at least $30\,\mev$.

Neutral pions are reconstructed by combining two photon candidates.
The invariant mass of a $\piz$ meson candidate is required to be within $[-20,\ +15]\,\mevcc$ of the nominal $\piz$ mass~\cite{PDG2016}.
The $\eta$ mesons are reconstructed in the decay modes $\eta \to \gamma\gamma$ and $\pip\pim\piz$.
The invariant mass is required to be within $[-25,\ +20]\,\mevcc$ and $\pm 10\,\mevcc$ of the nominal $\eta$ mass~\cite{PDG2016} for $\eta \to \gamma\gamma$ and  $\eta \to \pip\pim\piz$ candidates, respectively.
The $\omega$ mesons are reconstructed in the decay mode $\omega \to \pip\pim\piz$. The invariant mass of an $\omega$ meson candidate is required to be within $[-15,\ +10]\,\mevcc$ of the nominal $\omega$ mass~\cite{PDG2016}.

Neutral kaons are reconstructed in the decay mode $\KS \to \pip\pim$. The invariant mass of a $\KS$ meson candidate is required to be within $\pm15\,\mevcc$ of the nominal value~\cite{PDG2016}.
Standard selection requirements exploiting the displacement of the $\KS$ decay vertex from the $e^+ e^-$ interaction point (IP) described in Refs.~\cite{BELLEKshortselection,BaBar_D0h0_2007_twobody} are applied.

Neutral $\D$ mesons are reconstructed in the decay mode \DtoKSpipi. The invariant mass of a $\D$ meson candidate is required to be within $\pm 15\,\mevcc$ of the nominal value~\cite{PDG2016}.
Neutral $\Dstar$ mesons are reconstructed in the decay mode $\Dstar \to \D \piz$. To select $\Dstar$ mesons, the reconstructed mass difference of neutral $\Dstar$ and $\D$ meson candidates
is required to be within $\pm 2.5\,\mevcc$ of the nominal value~\cite{PDG2016}.

Neutral $\B$ mesons are reconstructed by combining light unflavored and neutral hadron candidates, $h^{0} \in \{\piz, \eta, \omega \}$, with $\D^{(*)}$ candidates.
The decay modes $\Bz \to \D \piz$, $\D \eta$, $\D \omega$, $\Dstar \piz$, and $\Dstar \eta$, where sufficient signal yields are reconstructed, are included in the analysis.
Neutral $\B$ mesons are selected using three variables that are constructed from kinematic observables:
the beam-energy-constrained mass \Mbcprime, the energy difference \DeltaE, and the neural network classifier \NNoutprime.

The beam-energy-constrained mass is defined as:
\begin{widetext}
\begin{equation}
      \Mbcprime = \sqrt{  {E^{*2}_{\rm beam}} / c^{4} - \left( \vec{p}^{*}_{D^{(*)}} / c + \frac{\vec{p}^{*}_{h^{0}}}{\lvert \vec{p}^{*}_{h^{0}} \rvert}  \sqrt{ \left( E^{*}_{\rm beam} - E^{*}_{D^{(*)}} \right)^{2} / c^{4} - M_{h^{0}}^{2} } \right)^{2} } \mathrm{,} \label{eqn_beam_energy_constrained_mass}   
\end{equation}
\end{widetext}
where $E^{*}_{\rm beam}$ is the energy of either beam provided by the $e^+ e^-$ collider,
the variables $\vec{p}^{*}_{D^{(*)}}$ and $E^{*}_{D^{(*)}}$ are the three-momentum and the energy of the $D^{(*)}$ meson candidates,
and $\vec{p}^{*}_{h^{0}}$ and $M_{h^{0}}$ are the three-momentum and the invariant mass of the $h^{0}$ candidates.
Observables marked with an asterisk are evaluated in the $e^+ e^-$ c.m.\ frame.
Belle introduced the variable \Mbcprime in the measurements of \B meson decays mediated by radiative penguin transitions~\cite{Mbcprime_definition}
as an alternative to the more commonly used variable $M_{\mathrm{bc}} = \sqrt{ {E^{*2}_{\rm beam}} / c^{4} - \vec{p}^{*2}_{\B} / c^{2} } = \sqrt{ {E^{*2}_{\rm beam} } / c^{4} - (\vec{p}^{*}_{D^{(*)}} + \vec{p}^{*}_{h^{0}})^{2} / c^{2} }$.
We note that \Mbcprime does not directly depend on the three-momentum magnitude nor the energy, but only on the direction of flight of the $h^{0}$ candidate.
Therefore, \Mbcprime is insensitive to potential correlations with the energy difference, defined as
\begin{equation}
   \DeltaE = E^{*}_{\B} - E^{*}_{\rm beam} \mathrm{.}
\end{equation}
In the present analysis, non-trivial correlations emerge between $M_{\mathrm{bc}}$ and \DeltaE for final states containing photons from the reconstructed $h^{0}$ decay modes
due to energy mismeasurements by the electromagnetic calorimeters, for example, caused by shower leakage effects.
The use of \Mbcprime effectively eliminates these correlations and enables factorizing the p.d.f.s constructed from the \Mbcprime and \DeltaE observables in multi-dimensional fits.

The neural network combines information characterizing the shape of the events and is based on 16 modified Fox-Wolfram moments~\cite{Neurobayes,FWmoments}.
Following an approach introduced by Belle in Ref.~\cite{NNoutprime_definition}, the variable \NNoutprime is constructed from the output of the neural network classifier, \NNout, by the following transformation:
\begin{equation}
  \NNoutprime = \log \frac{ \NNout - \NNoutmin }{ \NNoutmax - \NNout } \label{eqn:nbout_transformed} \mathrm{.}
\end{equation}
The variables \NNoutmin and \NNoutmax are adjustable parameters, and are related to the output domain of \NNout.
In this analysis, $\NNoutmin = 0.2 $ and $ \NNoutmax = 1$ are chosen.
After the transformation to \NNoutprime, the output of the neural network classifier exhibits smooth distributions around a peak position that differs for $e^{+}e^{-} \to q \overline{q}$  $( q \in \{ u, d, s, c \} )$ continuum events
and $B\kern 0.18em\overline{\kern -0.18em B}$ events.
Candidates from continuum events tend to be distributed around a peak position at negative values of \NNoutprime,
while $B\kern 0.18em\overline{\kern -0.18em B}$ events are distributed around a peak position at positive values.
The \NNoutprime distributions can be described by empirical parameterized models with few d.o.f., such as the
the Novosibirsk function, an empirical p.d.f.\  inspired by the log-normal distribution and defined in Ref.~\cite{NovosibriskFunction}.
The use of a parameterized model has technical advantages when including the neural network classifier in addition to \Mbcprime and \DeltaE in multi-dimensional fits
to extract the \BtoDhzero signal.
Before applying the transformation described above, a loose requirement of $\NNout > 0.2$ is applied to remove regions that are almost exclusively populated by continuum background events.

The following requirements are applied on \Mbcprime, \DeltaE, and \NNoutprime to select neutral $\B$ mesons:
$5.24 < \Mbcprime < 5.29\gevcc$,
$-150 < \DeltaE < 200\mev$, and
$-8 < \NNoutprime < 10$.

\subsection{Estimation of the {\boldmath$\Bz \to \D^{(*)} h^{0}$} signal yields}
\label{sec:Bdecay_determination_signal_yields}
The \BtoDhzero signal yields are determined by three-dimensional extended unbinned ML fits to the \Mbcprime, \DeltaE, and \NNoutprime distributions.
The fit model accounts for five components and is described below.

For \BtoDhzero signal decays, the \Mbcprime, \DeltaE, and \NNoutprime distributions exhibit smooth peaking structures.
The shapes of the signal component are parameterized by two Novosibirsk functions for \Mbcprime, one symmetric and two two-piece normal distributions for \DeltaE,
and two Novosibirsk functions for \NNoutprime. The signal shapes are calibrated using the high-statistics data control sample of $\Bz \to \Db^{(*)0} h^{0}$ decays with the CKM-favored $\Dzb \to \Kp \pim$ decay.

For $\Bz \to \D^{} h^{0}$ decays, candidates can originate from the corresponding $\Bz \to \D^{*} h^{0}$ decay modes, if the slow neutral pion from $\Dstar \to \D \piz$ decays is missed during the reconstruction.
This ``crossfeed component'' originates from true $\Bz \to \D^{*} h^{0}$ signal decays and has therefore signal-like properties.
The crossfeed has similar shapes as the signal but peaks at negative \DeltaE.
The contribution of the crossfeed is small, at the level of $3-13\%$ with respect to the signal. In the fits, the fractions of this component are fixed to the values estimated from high-statistics MC simulations of signal decays.
The shapes of the crossfeed component are parameterized by two Novosibirsk functions for \Mbcprime, one kernel density estimator for \DeltaE, and two Novosibirsk functions for \NNoutprime.

In addition to the contributions from the signal and the signal-like crossfeed, the fit model accounts for the following three separate sources of background.
The first source originates from partially-reconstructed $\Bp \to \overline{\kern -0.2em D}{}^{(*)0} \rho^{+}$ decays,
which constitute a background for $\Bz \to \D^{(*)} \piz$ decays when the charged pion from $\rho^{+} \to \pip \piz$ decays is soft.
This background arises only for $\Bz \to \D^{} \piz$ and $\Bz \to \D^{*} \piz$ decays, but is not present for the other \BtoDhzero decay modes.
Like the crossfeed component, the background from $\Bp \to \overline{\kern -0.2em D}{}^{(*)0} \rho^{+}$ decays has a similar shape as the signal, but peaks at negative \DeltaE.
The shapes are parameterized by two Novosibirsk functions for \Mbcprime, one kernel density estimator for \DeltaE, and two Novosibirsk functions for \NNoutprime.
The $\Bp \to \overline{\kern -0.2em D}{}^{(*)0} \rho^{+}$ background component is determined by the fit.

The second source of background arises from \B meson candidates formed from random combinations of final state particles originating from $e^{+}e^{-} \to B\kern 0.18em\overline{\kern -0.18em B}$ events.
This ``combinatorial $B\kern 0.18em\overline{\kern -0.18em B}$ background'' is low in the present analysis. The combinatorial $B\kern 0.18em\overline{\kern -0.18em B}$ background exhibits smooth phase space distributions in \Mbcprime and \DeltaE,
and peaks at positive \NNoutprime.
The shapes are parameterized by an ARGUS function~\cite{ARGUSfunction} for \Mbcprime, a second-order polynomial function for \DeltaE, and two Novosibirsk functions for \NNoutprime.

The third source of background originates from $e^{+}e^{-} \to q \overline{q}$  $( q \in \{ u, d, s, c \} )$ continuum events.
This continuum background exhibits smooth phase space distributions in \Mbcprime and \DeltaE, and peaks at negative \NNoutprime.
The shapes are parameterized by an ARGUS function for \Mbcprime, a second-order polynomial function for \DeltaE, and two Novosibirsk functions for \NNoutprime.

\begin{figure*}[htb]
\includegraphics[width=0.95\textwidth]{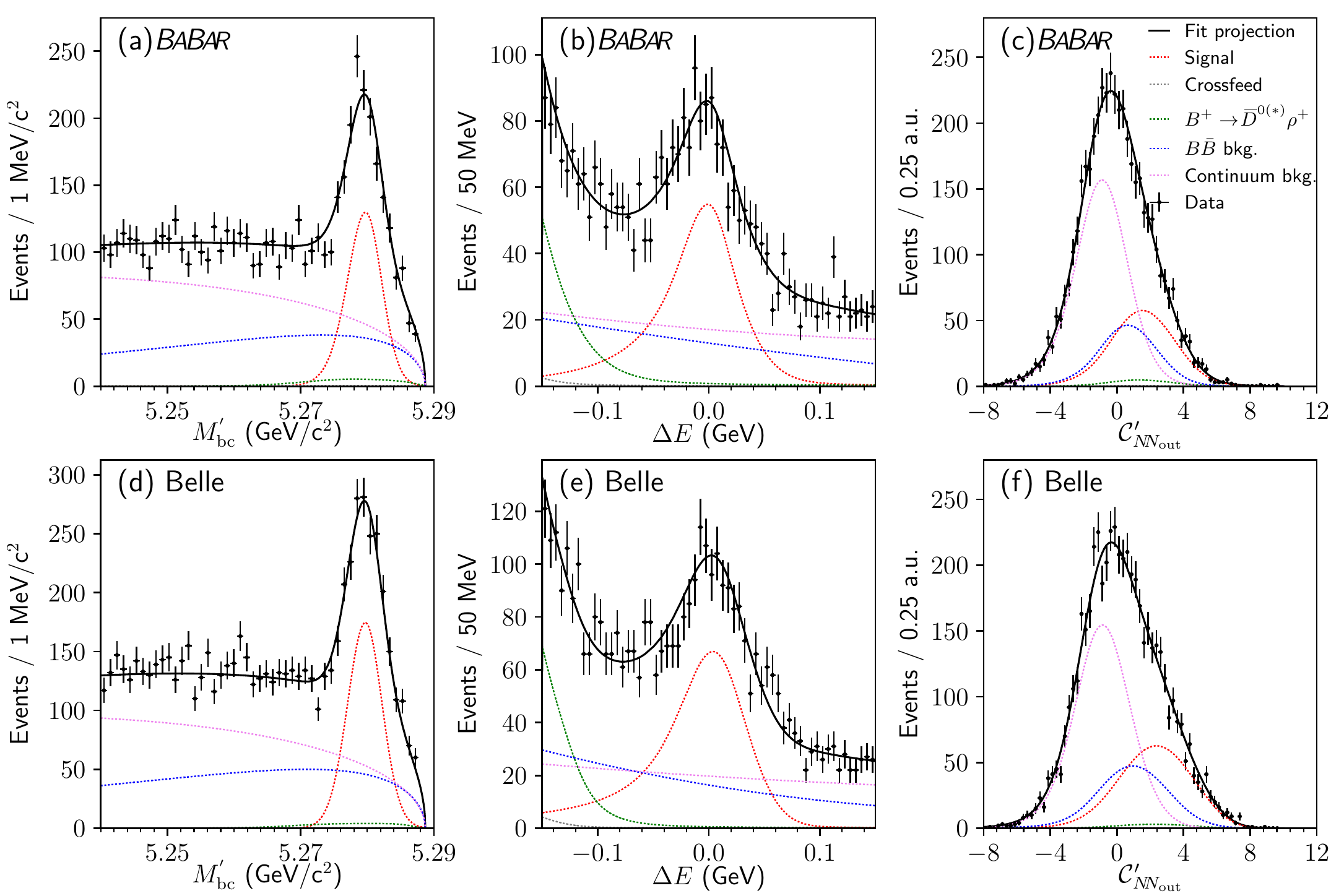}
\caption{
(color online). Data distributions for \Mbcprime (left), \DeltaE (center), and \NNoutprime (right) for \BtoDhzero decays (points with error bars) reconstructed from \babar\ (top) and Belle (bottom) data.
The solid black lines represent projections of the total fit function, and the colored dotted lines show the signal and background components of the fit as indicated in the upper-right panel's legend.
In plotting the \Mbcprime, \DeltaE, and \NNoutprime distributions, each of the other two observables are required to satisfy
$\Mbcprime > 5.272\gevcc$, $\lvert\Delta E\rvert<100\mev$, or $0 < \NNoutprime < 8$ to select signal-enhanced regions.
}
\label{figure_Mbc_DeltaE_nbout_fit}
\end{figure*}

In total, \BtoDhzero signal yields of $ 1\,129 \pm 48 $ events for \babar\ and $ 1\,567 \pm 56 $ events for Belle are obtained.
The signal yields separated by experiment and decay mode are summarized in Table~\ref{table:data_yields_decay_modes}.
The experimental \Mbcprime, \DeltaE, and \NNoutprime distributions and projections of the fits are shown in Fig.~\ref{figure_Mbc_DeltaE_nbout_fit}.

\begin{table}[htb]
\caption{Summary of the $\Bz \to D^{(*)} h^{0}$ signal yields determined by the three-dimensional extended unbinned ML fits to the \Mbcprime, \DeltaE, and \NNoutprime distributions described in Sect.~\ref{sec:Bdecay_determination_signal_yields}.}
\label{table:data_yields_decay_modes}
\begin{tabular}{l D{,}{\pm}{-1} D{,}{\pm}{-1} }
\hline \hline
\multicolumn{1}{l}{Decay mode} & \multicolumn{1}{c}{\babar} & \multicolumn{1}{c}{Belle} \\
\hline
$\Bz \to \D \piz$                      & 469,31 & 768,37 \\
$\Bz \to \D \eta$                      & 220,22 & 238,23 \\
$\Bz \to \D \omega$                    & 219,21 & 285,26 \\
$\Bz \to \D^{*}\piz$                  & 147,18 & 182,19 \\
$\Bz \to \D^{*}\eta$                  &  74,11 &  94,13 \\
\hline
Total & 1\,129,48 & 1\,567,56 \\
\hline \hline
\end{tabular}
\end{table}

%%%%%%%%%%%%%%%%%%%%%%%%%%%%%%%%%%%%%%%%%%%%%%%%%%%%%%%%%%%%%%%%%%%%%%%%%%%%%%%%%%%%%%%%%%%%%%%%%%%%%%%%%
%%% Figure: Mbc', DeltaE and NNout_transformed BELLE
%%%%%%%%%%%%%%%%%%%%%%%%%%%%%%%%%%%%%%%%%%%%%%%%%%%%%%%%%%%%%%%%%%%%%%%%%%%%%%%%%%%%%%%%%%%%%%%%%%%%%%%%%

\subsection{Time-dependent Dalitz plot analysis}
\label{subsection_timedependent_Dalitz_plot_analysis}
At \babar\ (Belle) the $\Upsilon(4S)$ is produced with a Lorentz boost of $\beta\gamma = 0.560$ ($0.425$), allowing the measurement of the proper-time interval between the decays of the two \B mesons.
The proper-time interval $\Delta t$ is given by $\Delta z / \mathrm{c} \beta \gamma$, where $\Delta z$ denotes the spatial distance between the decay vertices of the two \B mesons in the laboratory frame.
The \babar\ and Belle techniques to measure the flavor-tagged proper-time intervals of the \B mesons and to extract the \CP violation parameters are described in detail in Refs.~\cite{BaBar_btoccs,Belle_btoccs,BaBar_btoccs2002,BelleVertexResolution,BelleTaggingNIM,BFactoriesBook,SpringerBookRoehrken}.
The \BtoDhzero signal decay vertices are reconstructed by kinematic fits that include experimental knowledge of the IP position.
For \babar, the applied vertex reconstruction algorithm simultaneously includes the complete \B meson decay tree, including all secondary decays, in the kinematic fit.
For Belle, the vertex reconstruction is performed in an iterative bottom-up approach starting with the final state particles.
The decay vertex and the $b$-flavor content of the accompanying \B meson are estimated from the reconstructed decay products not assigned to the signal \B meson.
The $b$-flavor content is inferred by the flavor-tagging procedures described in Refs.~\cite{BaBar_btoccs,BelleTaggingNIM}. The applied algorithms account for different signatures such as the presence and properties of prompt leptons, charged kaons, and pions
originating from the decay of the accompanying \B meson, and assign a flavor and an associated probability.

The experimental conditions and the instrumentation of the detectors are different for \babar\ and Belle.
The finite experimental resolution in the measurements of proper-time intervals are different for \babar\ and Belle,
and both experiments follow different approaches to describe the resolution effects.
The two experiments employ different multivariate techniques for the flavor-tagging. \babar\ uses a neural network-based approach and Belle uses a multi-dimensional likelihood approach.

The time-dependent Dalitz plot analysis to measure the \CP violation parameters follows the technique established in the previous combined \babar+Belle time-dependent \CP violation measurement of $\Bzb \to D^{(*)}_{\CP} h^{0}$ decays~\cite{Roehrken2015}.
The strategy of the combined approach is to apply established, experiment-specific techniques to describe proper-time resolution and flavor-tagging effects by \babar\ and Belle to the data collected by the particular experiment.
The combined measurement is then performed by maximizing the log-likelihood function constructed from the p.d.f.s and the data collected by both experiments:
\begin{equation}
\ln \mathcal{L} = \sum \limits_{i} \ln \mathcal{P}_{i}^{\mathrm{\scriptsize\babar}} + \sum \limits_{j} \ln \mathcal{P}_{j}^{\mathrm{Belle}} \mathrm{.} \label{equation_loglikelihood}
\end{equation}
The indices $i$ and $j$ run over events reconstructed from \babar\ and Belle data, respectively.
All events used in the \Mbcprime, \DeltaE, and \NNoutprime fits are included.
The $\mathcal{P}$ are the p.d.f.s of the experimental flavor-tagged proper-time interval and Dalitz plot distributions of the \B mesons measured in the events, and are defined as:
\begin{align}
\mathcal{P} =
\sum\limits_{k}
f_{k} 
\int
\left[
{P}_{k} \left( \Delta t' \right)
R_{k} \left( \Delta t - \Delta t' \right) 
\right]
d\left( \Delta t' \right ) .
\end{align}
The index $k$ represents the signal and background components. The fractions of the components, $f_{k}$, are evaluated on an event-by-event basis as a function of \Mbcprime, \DeltaE, and \NNoutprime.
The ${P}_{k}$ are the p.d.f.s that describe the particular underlying particle physics process and are the same for both experiments.
The ${P}_{k}$ are convolved with the resolution functions $R_{k}$ that account for the finite proper-time resolution.

For the signal, the p.d.f.s are constructed from Eqs.~(\ref{equation:decay_rate}) and~(\ref{equation_sine_cosine}) convolved with the experiment-specific resolution functions to account for the finite proper-time resolution~\cite{BaBar_btoccs,BelleVertexResolution},
and include the effect of incorrect flavor assignments by the applied flavor-tagging algorithms~\cite{BaBar_btoccs,BelleTaggingNIM}
and a correction to account for the variations of the reconstruction efficiency as a function of the position on the Dalitz plot.

Neutral \D mesons produced in \BtoDhzero decays have a different momentum spectrum than those produced in ${e^{+}e^{-} \to c\bar{c}}$ events.
In addition, the yield for the \BtoDhzero decay modes studied by the combined \babar+Belle approach is about three orders of magnitude lower than that for the \DzerotoKSpipi decays reconstructed from ${e^{+}e^{-} \to c\bar{c}}$ events.
Therefore, the Dalitz plot reconstruction efficiency correction used for the analysis of \BtoDhzero decays is different from that described in Sect.~\ref{sec:D_decay_Dalitz_plot_reconstruction_efficiency_correction},
and a parametrization with fewer d.o.f. is chosen.
The reconstruction efficiency map is constructed separately for \babar\ and Belle by the fit of a two-dimensional $3^{\mathrm{rd}}$-order polynomial function
in the Dalitz plot variables \MsquaredKSpiRS and \MsquaredKSpiWS to the reconstruction efficiency distributions
obtained from high-statistics samples of MC events of \BtoDhzero with \DtoKSpipi signal decays.

For the signal-like crossfeed from partially-reconstructed $\Bz \to \D^{*} h^{0}$ decays, the p.d.f.s are constructed as for the signal, but account for distinct properties
such as the \CP-eigenvalues of the particular final states of the crossfeed contribution.
The charged \B meson background from partially-reconstructed $\Bp \to \overline{\kern -0.2em D}{}^{(*)0} \rho^{+}$ decays is parameterized by an exponential p.d.f.\  accounting for the \Bp lifetime convolved with the experiment-specific resolution functions.
The combinatorial $B\kern 0.18em\overline{\kern -0.18em B}$ background and the continuum background share the same parametrization for \babar\ and Belle.
For each background component, the p.d.f.s are constructed from the sum of a Dirac delta function to model background from prompt particles and an exponential p.d.f.\  with effective lifetimes to model the non-prompt background.
The background p.d.f.s are convolved with a resolution function modeled as the sum of two Gaussian functions whose widths depend linearly on the uncertainty of \Deltat.
The \Deltat parameters for the combinatorial $B\kern 0.18em\overline{\kern -0.18em B}$ background and the continuum background
are determined by fits to the $M_{\mathrm{bc}}^{\prime} < 5.26\,\gevcc$ sidebands and are fixed in the measurement.

In the fit, the parameters $\tau_{\Bz}$, $\tau_{\Bp}$, and $\Delta m_{d}$ are fixed to the world averages~\cite{HFAG}, and the Dalitz plot amplitude model parameters are fixed to the results
of the \DzerotoKSpipi Dalitz plot fit described above.
The only free parameters are $\sin{2\beta}$ and $\cos{2\beta}$, and the results are:
\begin{align}
\sin{2\beta} = 0.80 \pm 0.14  \,(\rm{stat.}) \pm 0.06 \,(\rm{syst.}) \pm 0.03 \,(\rm{model}) , \nonumber \\
\cos{2\beta} = 0.91 \pm 0.22  \,(\rm{stat.}) \pm 0.09 \,(\rm{syst.}) \pm 0.07 \,(\rm{model}) .
\end{align}
The linear correlation between $\sin{2\beta}$ and $\cos{2\beta}$ is $5.1\%$.
The result deviates less than $1.0$ standard deviations from the trigonometric constraint given by $\sin^{2}{2\beta} + \cos^{2}{2\beta} = 1$.

An alternative fit is performed to measure directly the \CP-violating phase $\beta$ using the signal p.d.f.\  constructed from Eq.~(\ref{equation:decay_rate}). The result of this fit is:
\begin{align}
\beta = \left( 22.5 \pm 4.4  \,(\rm{stat.}) \pm 1.2 \,(\rm{syst.}) \pm 0.6 \,(\rm{model}) \right)^{\circ}
\end{align}
The evaluation of the experimental systematic uncertainties and the uncertainties due to the applied \DzerotoKSpipi decay amplitude model are described
in Sects.~\ref{sec:experimental_systematic_uncertainties} and~\ref{sec:uncertainties_due_to_the_Dalitz_plot_amplitude_model}.

\begin{figure*}
\includegraphics[width=0.9\textwidth]{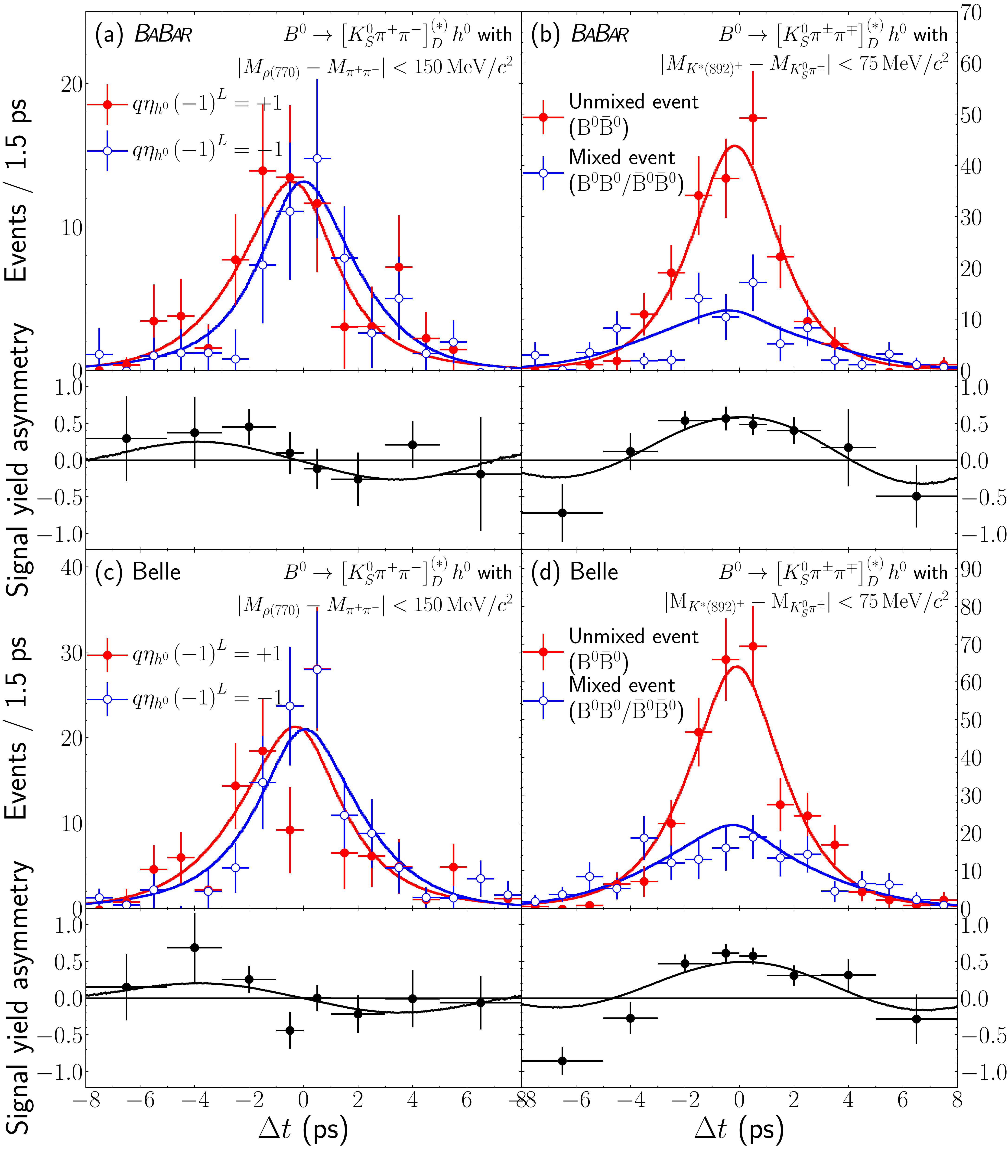}
\caption{
(color online). Distributions of the proper-time interval (data points with error bars) and the corresponding asymmetries for \BtoDhzero candidates associated
with high-quality flavor tags (\babar: lepton or kaon tagging categories; Belle: $r > 0.5$) for the \babar\ (top) and Belle (bottom) data samples.
The background has been subtracted using the \sPlot technique~\cite{sPlotNIM} with weights obtained from the
fit presented in Fig.~\ref{figure_Mbc_DeltaE_nbout_fit}.
Two different regions of the \DtoKSpipi phase space are shown.
In the plots of the left column, a region predominantly populated by \CP eigenstates,
$\Bz \to \left[ \KS \rho(770)^{0} \right]^{(*)}_{D} h^{0}$,
is selected by requiring
$\lvert M_{\rho(770)} - M_{\pip \pim} \rvert < 150\,\mevcc$.
In the plots of the right column, a region predominantly populated by quasi-flavor-specific decays,
$\Bz \to \left[ K^{*}(892)^{\pm} \pi^{\mp} \right]^{(*)}_{D} h^{0}$,
is selected by requiring
$\lvert M_{K^{*}(892)^{\pm}} - M_{\KS \pi^{\pm}} \rvert < 75\,\mevcc$.
}
\label{figure:DeltaT_data_distributions_and_projections_of_the_fit}
\end{figure*}

The $\Bz \to \left[ \KS \pip \pim \right]^{(*)}_{D} h^{0}$ decays proceeds via a rich variety of intermediate \CP eigenstates and quasi-flavor-specific decays contributing to the multibody final state.
These intermediate contributions involve different physics in the time evolution of the \B meson decay, and hence exhibit different proper-time interval distributions.
In Fig.~\ref{figure:DeltaT_data_distributions_and_projections_of_the_fit}, the proper-time interval distributions and projections of the fit for $\sin{2\beta}$ and $\cos{2\beta}$ are shown for two different regions of the \DzerotoKSpipi phase space.

In Figs.~\ref{figure:DeltaT_data_distributions_and_projections_of_the_fit}a and c, 
a region of phase space predominantly populated by \CP eigenstates, $\Bz \to \left[ \KS \rho(770)^{0} \right]^{(*)}_{D} h^{0}$, is selected by requiring $\lvert M_{\rho(770)} - M_{\pip \pim} \rvert < 150\,\mevcc$.
Since the $\left[ \KS \rho(770)^{0} \right]^{(*)}_{D} h^{0}$ final state is accessible for both \Bz and \Bzb,
interference between the amplitude for direct decays of neutral \B mesons into this final state and that following \Bz-\Bzb oscillations emerges.
The time evolution exhibits time-dependent \CP violation governed by the \CP-violating weak phase $2\beta$.
The proper-time interval distributions show the characteristic pattern for mixing-induced \CP violation, and the corresponding
time-dependent \CP asymmetry follows a sine oscillation similiar to our previous combined \babar+Belle measurement of $\sin{2\beta}$ in $\Bzb \to D^{(*)}_{\CP} h^{0}$ decays
with $D_{\CP}$ decaying into two-body \CP eigenstates~\cite{Roehrken2015}.

In Figs.~\ref{figure:DeltaT_data_distributions_and_projections_of_the_fit}b and d, 
regions of phase space predominantly populated by quasi-flavor-specific decays, $\Bz \to \left[ K^{*}(892)^{\pm} \pi^{\mp} \right]^{(*)}_{D} h^{0}$, are selected by requiring $\lvert M_{K^{*}(892)^{\pm}} - M_{\KS \pi^{\pm}} \rvert < 75\,\mevcc$.
The decays of neutral \B mesons to the $\left[ K^{*}(892)^{\pm} \pi^{\mp} \right]^{(*)}_{D} h^{0}$ final states are, to a good approximation, flavor-specific.
Therefore, no interference between \Bz and \Bzb mesons and no time-dependent \CP violation can emerge.
Instead, the time evolution exhibits \Bz-\Bzb oscillations governed by the decay width difference of the physical eigenstates of neutral \B mesons (\Bz-\Bzb oscillation frequency), $\Delta m_{d}$.
The proper-time interval distributions show the characteristic oscillation pattern for quantum-entangled \B meson pairs produced and tagged in $e^+ e^- \to \Upsilon\left(4S\right) \to \Bz\Bzb$ events.
The Einstein-Podolsky-Rosen (EPR) effect~\cite{EinsteinPodolskyRosen1935} prevents the two neutral \B mesons from being produced with the same flavor at $\Delta t = 0$, which in Figs.~\ref{figure:DeltaT_data_distributions_and_projections_of_the_fit}b and d
is additionally smeared by experimental resolution effects.
The time evolution follows a $1 \pm \cos(\Delta m \Delta t)$ distribution, and the corresponding time-dependent oscillation asymmetry exhibits a cosine oscillation.

Various cross-checks are performed to validate the procedure of the measurement.
The $\Bz \to \Db^{(*)0} h^{0}$ decays with the CKM-favored $\Dzb \to \Kp \pim$ decay have very similar kinematics and background composition as \BtoDhzero with \DtoKSpipi decays and provide a high-statistics control sample.
In total, signal yields of $ 3\,029 \pm 73 $ events for \babar\ and $ 4\,042 \pm 84 $ events for Belle are obtained for the control sample.
Using the same analysis approach, the time-dependent \CP violation measurement of the control sample yields both mixing-induced and direct \CP violation consistent with zero, in agreement with the expectation of negligible \CP violation for these flavor-specific decays. 
Measurements of the neutral \B meson lifetime for \BtoDhzero with \DtoKSpipi decays and for the control sample without flavor-tagging applied yield
$\tau_{\Bz} = \left(1.500 \pm 0.052\,(\rm{stat.})\right)\,\mathrm{ps}$ and $\tau_{\Bz} = \left(1.535 \pm 0.028\,(\rm{stat.})\right)\,\mathrm{ps}$, respectively, and are in agreement with the world average $\tau_{\Bz} = \left(1.520 \pm 0.004\right)\,\mathrm{ps}$~\cite{HFAG}.
In addition, all measurements have been performed for data separated by experiments and yield consistent results.
The results for \BtoDhzero with \DtoKSpipi decays separated by experiments are
$\sin{2\beta} = 0.91 \pm 0.20\,(\rm{stat.})$,
$\cos{2\beta} = 0.87 \pm 0.31\,(\rm{stat.})$,
and
$\beta = \left( 25.6 \pm 6.4  \,(\rm{stat.}) \right)^{\circ}$ for \babar,
and
$\sin{2\beta} = 0.70 \pm 0.20\,(\rm{stat.})$,
$\cos{2\beta} = 0.96 \pm 0.30\,(\rm{stat.})$,
and
$\beta = \left( 19.6 \pm 6.1  \,(\rm{stat.}) \right)^{\circ}$ for Belle, respectively.

%%%%%%%%%%%%%%%%%%%%%%%%%%%%%%%%%%%%%%%%%%%%%%%%%%%%%%%%%%%%%%%%%%%%%%%%%%%%%%%%%%%%%%%%%%%%%%%%%%%%%%%%%
%%% Figure: DeltaT data distributions and projections of the fit
%%%%%%%%%%%%%%%%%%%%%%%%%%%%%%%%%%%%%%%%%%%%%%%%%%%%%%%%%%%%%%%%%%%%%%%%%%%%%%%%%%%%%%%%%%%%%%%%%%%%%%%%%

\subsection{Determination of the systematic uncertainties}
The present analysis accounts for two classes of systematic uncertainties on the measured \CP violation parameters:
first, the experimental systematic uncertainty accounts for experimental effects that can affect the time-dependent Dalitz plot analysis;
second, the Dalitz plot model uncertainty accounts for assumptions made on the applied \DzerotoKSpipi decay amplitude model.

\subsubsection{Experimental systematic uncertainties}
\label{sec:experimental_systematic_uncertainties}
The estimation of the experimental systematic uncertainty on the \CP violation parameters follows established methods, described in Refs.~\cite{BaBar_btoccs,Belle_btoccs,Roehrken2015}.
The evaluation of the individual contributions to the experimental systematic uncertainty are described below, and the results are summarized in Table~\ref{systemtic_uncertainties_cpfit}.

The systematic uncertainty due to vertex reconstruction accounts for the
applied vertex reconstruction algorithms, the requirements applied to select \B mesons,
the uncertainty of the $z$ scale, possible $\Delta t$ biases, and effects due to possible
misalignment of the silicon vertex detectors.
For the vertex reconstruction algorithms, the constraints in the kinematic fits and
the applied selection requirements of the signal \B meson and the accompanying \B meson are varied.
For \babar, the uncertainty due to the $z$ scale and the Lorentz boost is estimated by variations of the
corresponding scale and uncertainties.
For Belle, a possible $\Delta t$ bias is estimated using MC simulations.
Misalignment effects of the silicon vertex detectors are studied by MC simulations, and corresponding
systematic uncertainties are assigned.

Experiment-specific resolution models are applied to account for effects due to the finite experimental \Deltat resolution.
The \Deltat resolution function parameters are fixed to values obtained from control samples using \babar\ and Belle data.
The systematic uncertainty due to the applied \Deltat resolution functions is estimated by variation of the resolution model parameters within their uncertainties.

The parameters of the \Deltat model for the combinatorial $B\kern 0.18em\overline{\kern -0.18em B}$ background and the continuum background
are determined by fits to the $\Mbcprime < 5.26\,\gevcc$ data sidebands.
The systematic uncertainty due to the background \Deltat p.d.f.s is estimated by variation of the \Deltat background model parameters within their uncertainties.

The signal purity is estimated by the three-dimensional unbinned ML fit to the \Mbcprime, \DeltaE, and \NNoutprime distributions.
The systematic uncertainty due to the signal purity estimation is estimated by variation of the fit parameters within their uncertainties.

The $b$-flavor content of neutral \B mesons is inferred by multivariate \babar- and Belle-specific flavor-tagging algorithms.
The flavor-tagging algorithms are calibrated using control samples reconstructed from \babar\ and Belle data.
The systematic uncertainty due to the flavor-tagging is estimated by variation of the wrong-tag fractions and
the corresponding wrong-tag fraction differences for each tagging category within their uncertainties.

The neutral \B lifetime \tauBzero, the charged \B meson lifetime \tauBplus, and the \Bz-\Bzb oscillation frequency \Deltam are fixed to the world averages.
The systematic uncertainty due to these fixed physics parameters is estimated by variation of the lifetimes and oscillation frequency within their uncertainties.

The systematic uncertainty due a possible small fit bias in \Deltat measurements is estimated by MC simulations.
Large MC samples are generated using a complex \DzerotoKSpipi decay amplitude model and with \CP violation,
the same reconstruction algorithms and event selection requirements are applied to the MC samples as for the data, and the time-dependent Dalitz plot analysis is performed.
The deviations of the central values of the \CP violation parameters measured using the MC samples from the nominal result are assigned as systematic uncertainties.

The effect due to the applied Dalitz plot reconstruction efficiency correction for neutral \D mesons produced in \BtoDhzero decays is estimated by removing the efficiency correction.
The time-dependent Dalitz plot analysis is performed without the efficiency correction, and assigning the deviations from the nominal result as systematic uncertainty due to the
Dalitz plot reconstruction efficiency correction.

Most systematic uncertainties are independent for \babar\ and Belle. Possible correlations such as for the physics parameters are considered.
Additional contributions to the systematic uncertainty from possible sources of peaking background and the tag-side interference have been considered and can be neglected in the presented measurement.

The total experimental systematic uncertainty is the quadratic sum of all contributions.

\begin{table*}[htb]
\caption{Experimental systematic uncertainties on the \CP violation parameters.}
\label{systemtic_uncertainties_cpfit}
\begin{tabular}
 {@{\hspace{0.5cm}}l@{\hspace{0.5cm}}  @{\hspace{0.5cm}}c@{\hspace{0.5cm}}  @{\hspace{0.5cm}}c@{\hspace{0.5cm}}  @{\hspace{0.5cm}}c@{\hspace{0.5cm}}}
\hline \hline
Source & $\delta\sin{2\beta}$ ($\times 10^{2}$) & $\delta\cos{2\beta}$ ($\times 10^{2}$) & $\delta\beta$ ($^\circ$) \\
\hline
Vertex reconstruction                                 &   $3.2$    &   $4.8$   &   $0.53$   \\
$\Delta t$ resolution functions                       &   $2.8$    &   $5.8$   &   $0.41$   \\
Background $\Delta t$ p.d.f.s                         &   $1.2$    &   $1.8$   &   $0.16$   \\
Signal purity                                         &   $2.1$    &   $3.4$   &   $0.53$   \\
Flavor-tagging                                        &   $0.3$    &   $0.4$   &   $0.07$   \\
Physics parameters                                    &   $0.1$    &   $0.1$   &   $0.02$   \\
Possible fit bias                                     &   $3.7$    &   $3.9$   &   $0.79$   \\
Dalitz plot reconstruction efficiency correction      &  $<0.1$    &   $0.2$   &   $0.02$   \\
\hline
Total                                                 &   $6.1$    &   $9.3$   &   $1.18$   \\
\hline \hline
\end{tabular}
\end{table*}

\subsubsection{Uncertainty due to the Dalitz plot amplitude model}
\label{sec:uncertainties_due_to_the_Dalitz_plot_amplitude_model}
The model uncertainty accounts for the dependence of the \CP violation parameters on the
\DzerotoKSpipi decay amplitude model determined by the Dalitz plot amplitude analysis using the high-statistics Belle $e^+ e^- \to c\bar{c}$
data sample described in Sect.~\ref{sec:D_decay_Dalitz_amplitude_analysis}.
The strategy to estimate the model uncertainty is to repeat the \DzerotoKSpipi Dalitz plot amplitude analysis
with alternative assumptions and variations of the \DzerotoKSpipi decay amplitude model.
The time-dependent Dalitz plot analysis of \BtoDhzero decays is then performed using the alternative models as input,
and the deviations from the result using the nominal \DzerotoKSpipi decay amplitude model are assigned
as model uncertainty on the \CP violation parameters.
The evaluation of the individual contributions to the uncertainty due to the Dalitz plot amplitude model are described below, and the results are summarized in Table~\ref{table:cpfit_systematics_Dalitz_model}.

For the masses and widths of resonances fixed to the world averages, each resonance parameter is varied within its uncertainty to estimate the associated model uncertainty.

The model uncertainty due to the chosen $\pip\pim$ $S$-wave parametrization using the $K$-matrix formalism is estimated
by replacing the nominal $K$-matrix solution by alternative solutions from Ref.~\cite{Kmatrix2003}.
In addition, the parameter $s_{0}^{\mathrm{prod}}$ is varied within its uncertainty, which is taken from Ref.~\cite{BABAR2008}.

The LASS parametrization is used to model the $K\pi$ $S$-waves. The model uncertainty is estimated by replacing the LASS parametrization for the $K^{*}_{0}(1430)^-$ and $K^{*}_{0}(1430)^+$ resonances by standard relativistic BW terms.

The model uncertainty due to the chosen Blatt-Weisskopf barrier factors for \D mesons and intermediate resonances is estimated by varying the fixed parameters $d_{\D}$ and $d_{r}$ each by $\pm 0.5 \, \hbar c / \mathrm{GeV}$.

The fraction of wrong \D meson flavor-tags of the flavor-tagged $c\bar{c}$ data sample is fixed to the value estimated from the fit to the \deltaMDstarDzero sideband region on data.
The \D meson mistag fraction is varied within its uncertainty to evaluate the associated model uncertainty.

The model uncertainty due the applied Dalitz plot reconstruction efficiency correction is estimated by replacing the parameterized efficiency map by the corresponding two-dimensional binned distributions.

In the Dalitz plot amplitude analysis, the background is described by a parameterized model taken from the \deltaMDstarDzero and \MDzero sideband regions on data.
The model uncertainty due to the applied background description is estimated by replacing the parameterized background model by the two-dimensional binned distributions from the data sidebands.

Most intermediate two-body resonances contributing to \DzerotoKSpipi decays have a natural width much larger than the finite experimental resolution of reconstructed invariant masses, and resolution effects can be neglected in the \DtoKSpipi Dalitz plot amplitude analysis.
The $\omega(782)$ width, $8.5\,\mathrm{MeV}$, is comparable to the mass resolution. To estimate the size of possible effects due to the mass resolution and to evaluate the associated model error, the width of the $\omega(782)$ is increased by $20\%$.

The signal and background fractions used in the Dalitz plot amplitude analysis are determined by the fit of the two-dimensional \deltaMDstarDzero and \MDzero distributions.
The model uncertainty due to the signal purity estimation is determined by varying the the \deltaMDstarDzero-\MDzero model parameters within their uncertainties.

The statistical uncertainties on the Dalitz plot amplitude model parameters that are summarized in Table~\ref{Dalitz_amplitude_model} are caused by the finite size of the $c\bar{c}$ data sample.
To propagate the statistical uncertainties to the \CP violation parameters and assign the associated model error, each parameter is varied within its uncertainty.
For individual resonances, the correlations between phases and amplitudes are accounted for. An explicit treatment of additional correlations between resonances important in the \CP violation measurement were found to be negligible.
The chosen approach has to be found sufficient given that this systematic uncertainty does not limit the precision of the measurement.

The dependence of the model on resonances with very small contributions is estimated by removing resonances with fit fractions
of $0.1\%$ or lower. The doubly Cabibbo-suppressed $K^{*}(1410)^{+}$, $K^{*}_{2}(1430)^{+}$, and $K^{*}_{0}(1430)^{+}$,
and the $K^{*}(1410)^{-}$ are each removed from the model. For each model variation, the \DzerotoKSpipi Dalitz plot amplitude analysis is repeated to estimate the associated model uncertainty.

As a further cross-check and estimate of the possible model-dependence, a pure isobar \DzerotoKSpipi decay model without
the $K$-matrix parametrization is constructed. As in the isobar model discussed in Sect.~\ref{sec:D_decay_Dalitz_fit_validation},
the intermediate resonant contributions to the $\pip\pim$ $S$-wave are modeled by the $\sigma_{1}$, $\sigma_{2}$, $f_{0}(980)$, and $f_{0}(1370)$ resonances,
and a term constant in phase space is included to account for nonresonant contributions.
The \DzerotoKSpipi Dalitz plot amplitude analysis and the time-dependent Dalitz plot analysis of \BtoDhzero decays are repeated using the alternative model,
and the deviations of the \CP violation parameters from the baseline result are assigned as model uncertainty.
The result with the isobar model agrees well with the baseline result, which indicates small overall model dependence and robustness of the measurement.

The total model uncertainty is the quadratic sum of all contributions. Overall, the uncertainty due to the Dalitz plot amplitude model is small compared to the statistical uncertainty and the experimental systematic uncertainty.

\begin{table*}[htb]
\caption{Uncertainties on the \CP violation parameters due to the Dalitz plot amplitude model.}
\label{table:cpfit_systematics_Dalitz_model}
\begin{tabular}
 {@{\hspace{0.5cm}}l@{\hspace{0.5cm}}  @{\hspace{0.5cm}}c@{\hspace{0.5cm}}  @{\hspace{0.5cm}}c@{\hspace{0.5cm}}  @{\hspace{0.5cm}}c@{\hspace{0.5cm}}}
\hline \hline
Source & $\delta\sin{2\beta}$ ($\times 10^{2}$) & $\delta\cos{2\beta}$ ($\times 10^{2}$) & $\delta\beta$ ($^\circ$) \\
\hline
Masses and widths of resonances                       &   $0.7$    &   $1.7$   &   $0.13$   \\  
$\pip\pim$ $S$-wave parametrization                   &   $1.1$    &   $1.9$   &   $0.11$   \\
$K\pi$ $S$-wave parametrization                       &   $1.0$    &   $1.6$   &   $0.38$   \\ 
Blatt-Weisskopf barrier factors                       &   $1.2$    &   $1.7$   &   $0.19$   \\   
$D$ meson mistag fraction                             &   $0.2$    &  $<0.1$   &   $0.04$   \\   
Dalitz plot reconstruction efficiency                 &   $0.9$    &   $0.9$   &   $0.06$   \\   
Dalitz plot background shape                          &  $<0.1$    &   $0.2$   &   $0.01$   \\ 
Effect of finite experimental mass resolution         &   $0.1$    &   $0.2$   &   $<0.01$  \\   
Signal purity                                         &  $<0.1$    &  $<0.1$   &   $0.01$   \\ 
Statistical uncertainties on resonance parameters     &   $1.6$    &   $5.0$   &   $0.37$   \\   
Removal of resonances                                 &   $0.6$    &   $1.3$   &   $0.09$   \\
Alternative isobar Dalitz plot model                  &   $0.7$    &   $2.8$   &   $0.08$   \\  
\hline
Total                                                 &   $2.9$    &   $6.9$   &   $0.61$   \\
\hline \hline
\end{tabular}
\end{table*}

%%%%%%%%%%%%%%%%%%%%%%%%%%%%%%%%%%%%%%%%%%%%%%%%%%%%%%%%%%%%%%%%%%%%%%%%%%%%%%%%%%%%%%%%%%%%%%%%%%%%%%%%%
%%% Figure: DeltaLogLikelihood
%%%%%%%%%%%%%%%%%%%%%%%%%%%%%%%%%%%%%%%%%%%%%%%%%%%%%%%%%%%%%%%%%%%%%%%%%%%%%%%%%%%%%%%%%%%%%%%%%%%%%%%%%
\begin{figure}[htb]
\includegraphics[width=0.42\textwidth]{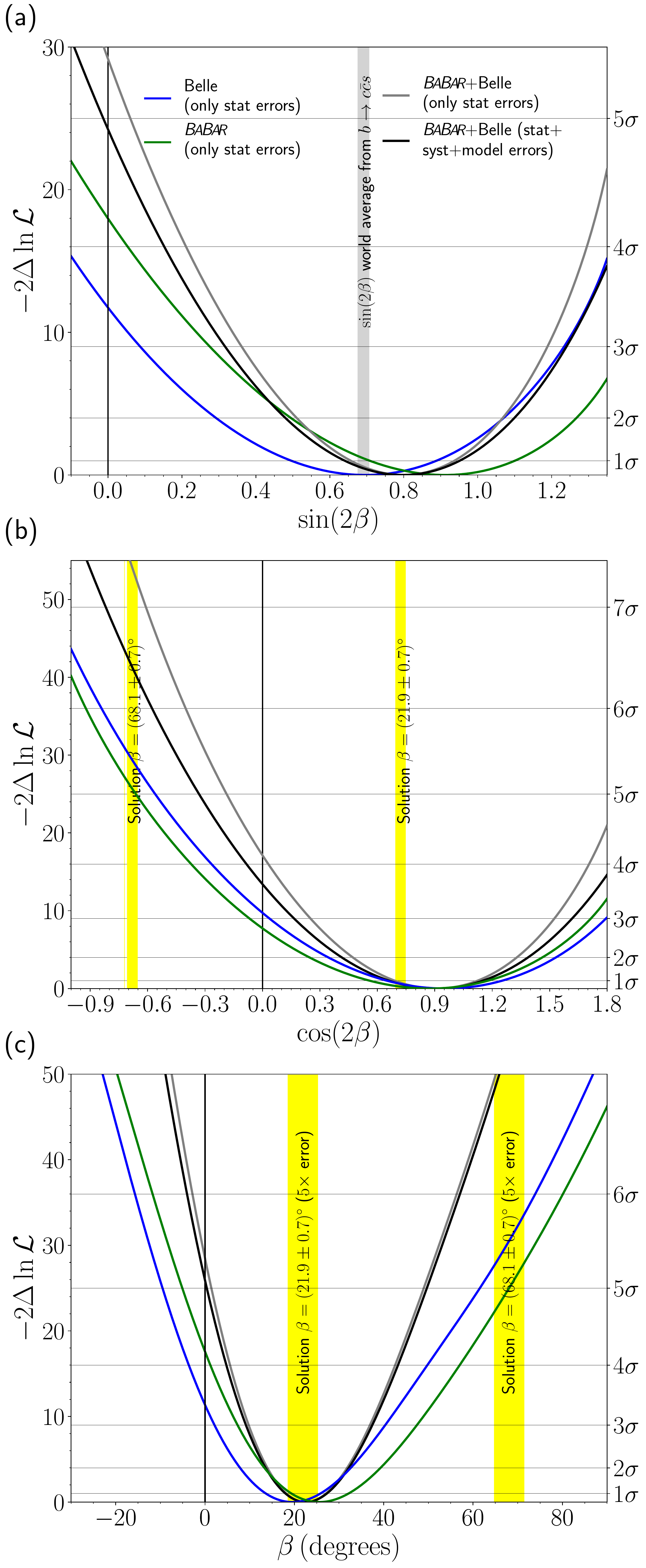}
\caption{
Obtained $-2 \Delta \mathrm{ln} \mathcal{L}$ curves for $\sin{2\beta}$, $\cos{2\beta}$, and $\beta$.
The black lines represent the results of the measurement including experimental systematic uncertainties and uncertainties due to the Dalitz plot amplitude model.
The green and blue lines represent the result of the measurement including only statistical uncertainties when using only \babar\ and Belle data, respectively.
}
\label{figure_likelihood_scan}
\end{figure}

\section{Interpretation of the results}
\label{sec:Bdecay_interpretation_of_the_results}
The statistical significance of the results is determined by a likelihood-ratio approach.
The change in $2\ln \mathcal{L}$ is computed when the \CP violation parameters are fixed to zero.
The experimental systematic uncertainties and the Dalitz plot amplitude model uncertainties are included by convolution of the likelihood curves.
The $-2 \Delta \mathrm{ln} \mathcal{L}$ curves for $\sin{2\beta}$ and $\cos{2\beta}$, and $\beta$ are shown in Fig.~\ref{figure_likelihood_scan}.
When computing $-2 \Delta \mathrm{ln} \mathcal{L}$ values for $\sin{2\beta}$ and $\cos{2\beta}$, the other observable is fixed to the nominal result.
The result for $\sin{2\beta}$ agrees within $0.7$ standard deviations with the world average of $\sin{2\beta} = 0.691 \pm 0.017$~\cite{HFAG} measured from $\bar{b} \to \bar{c}c\bar{s}$ transitions.
The measurement excludes the hypothesis of $\cos{2\beta}\leq 0$ at a $p$-value of $2.5 \times 10^{-4}$.
This corresponds to a significance of $3.7$ standard deviations, and thus provides the first evidence for $\cos{2\beta}>0$.
The results exclude the hypothesis of $\beta=0^{\circ}$ at a $p$-value of $3.6 \times 10^{-7}$.
This corresponds to a significance of $5.1$ standard deviations, and thus to an observation of \CP violation in $\Bzb \to D^{(*)} h^{0}$ decays.
The measured value for $\beta$ is in very good agreement with the preferred solution of the Unitarity Triangle with the world average of $(21.9 \pm 0.7)^{\circ}$~\cite{HFAG}.
The second solution of $\pi/2 - \beta = (68.1 \pm 0.7)^{\circ}$ is excluded with a $p$-value of $2.31 \times 10^{-13}$, corresponding to a significance of $7.3$ standard deviations.
Therefore, the present measurement reduces an ambiguity in the determination of the parameters of the CKM Unitarity Triangle.

\section{Conclusion}
\label{sec:conclusion}
In summary, we have measured $\sin{2\beta}$ and $\cos{2\beta}$ with a time-dependent Dalitz plot analysis of \BtoDhzero with \DtoKSpipi decays.
The analysis introduces several improvements over previous related measurements, and new concepts.
First, the measurement is performed by a simultaneous analysis of the final data samples collected by the \babar\ and Belle experiments,
totaling about $1.1\,\mathrm{ab}^{-1}$ and containing about $1\,240 \times 10^{6}$ $\B\Bb$ pairs collected at the $\Upsilon\left(4S\right)$ resonance .
The novel combined approach enables the doubling of the statistics available for the measurement, and allows the application of common assumptions
and the same \DzerotoKSpipi decay amplitude model simultaneously to the data collected by both experiments.
Second, a full Dalitz plot amplitude analysis is performed to derive the \DzerotoKSpipi decay amplitude model directly from a high-statistics $e^{+}e^{-} \to c\bar{c}$ data sample.
This enables full control over the model-building process, and the propagation of the \DzerotoKSpipi decay amplitude model uncertainties to those of the \CP violation parameters.
These approaches lead to improvements in the experimental sensitivity and in the robustness of the measurement.

We measure $\sin{2\beta} = 0.80 \pm 0.14  \,(\rm{stat.}) \pm 0.06 \,(\rm{syst.}) \pm 0.03 \,(\rm{model})$, $\cos{2\beta} = 0.91 \pm 0.22  \,(\rm{stat.}) \pm 0.09 \,(\rm{syst.}) \pm 0.07 \,(\rm{model})$, and
$\beta = \left( 22.5 \pm 4.4  \,(\rm{stat.}) \pm 1.2 \,(\rm{syst.}) \pm 0.6 \,(\rm{model}) \right)^{\circ}$.
The results on $\sin{2\beta}$ agree well with more precise measurements of $\bar{b} \to \bar{c}c\bar{s}$ transitions, and with our previous measurement combining \babar\ and Belle data~\cite{Roehrken2015}.
We exclude the hypothesis of $\beta=0^{\circ}$ at a significance of $5.1$ standard deviations, and we report an observation of \CP violation in $\Bzb \to D^{(*)} h^{0}$ decays.
We report the world's most precise measurement of the cosine of the \CP-violating weak phase $2 \beta$
and obtain the first evidence for $\cos{2\beta}>0$ at the level of $3.7$ standard deviations.
The measurement directly excludes the trigonometric multifold solution of $\pi/2 - \beta = (68.1 \pm 0.7)^{\circ}$ without further assumptions,
and thus resolves an ambiguity in the determination of the apex of the CKM Unitarity Triangle.

Moreover, the \BtoDhzero decays allow a theoretically cleaner determination of the \CP-violating phase $2\beta$ than the ``gold plated'' decay modes mediated by $\bar{b} \to \bar{c}c\bar{s}$ transitions~\cite{Fleischer2003}.
Therefore, future more precise measurements of \BtoDhzero decays can provide a new and complementary SM reference for $2\beta$. 

The combined \babar+Belle approach allows the access to an unprecedented large data sample
totaling more than $1\,\mathrm{ab}^{-1}$ recorded at c.m.\ energies of the $\Upsilon(4S)$ resonance
and enables a unique experimental precision, in particular, for time-dependent \CP violation measurements in the neutral \B meson system.
Our results underline the importance and discovery potential of future heavy flavor physics experiments
operated at high instantaneous luminosity such as the \B factory experiment Belle II~\cite{BelleIITDR},
which is expected to collect a data sample of $1\,\mathrm{ab}^{-1}$ by the year 2020
and is designed to collect $50\,\mathrm{ab}^{-1}$ by the middle of the next decade.

\section*{Acknowledgments}
\label{sec:acknowledgments}
We thank the \pep2\ and KEKB groups for the excellent operation of the accelerators.
The \babar\ experiment acknowledges the expertise and dedication of the computing organizations for their support.
The collaborating institutions wish to thank SLAC for its support and the kind hospitality extended to them. 
The Belle experiment wishes to acknowledge the KEK cryogenics group for the efficient
operation of the solenoid; and the KEK computer group,
the National Institute of Informatics, and the 
PNNL/EMSL computing group for valuable computing
and SINET5 network support.
This work was supported by the
the Ministry of Education, Culture, Sports, Science, and
Technology (MEXT) of Japan, the Japan Society for the 
Promotion of Science (JSPS), and the Tau-Lepton Physics 
Research Center of Nagoya University; 
the Australian Research Council;
Austrian Science Fund under Grant No.~P 26794-N20;
the Canadian Natural Sciences and Engineering Research Council;
the National Natural Science Foundation of China under Contracts 
No.~10575109, No.~10775142, No.~10875115, No.~11175187, No.~11475187, 
No.~11521505 and No.~11575017;
the Chinese Academy of Science Center for Excellence in Particle Physics; 
the Ministry of Education, Youth and Sports of the Czech
Republic under Contract No.~LTT17020;
the French Commissariat \`a l'Energie Atomique and Institut National de Physique Nucl\'eaire et de Physique des Particules;
the German Bundesministerium f\"ur Bildung und Forschung,
the Carl Zeiss Foundation, the Deutsche Forschungsgemeinschaft,
the Excellence Cluster Universe, and the VolkswagenStiftung;
the Department of Science and Technology of India; 
the Istituto Nazionale di Fisica Nucleare of Italy; 
National Research Foundation (NRF) of Korea Grants No. 2014R1A2A2A01005286, No. 2015R1A2A2A01003280,
No. 2015H1A2A1033649, No. 2016R1D1A1B01010135, No. 2016K1A3A7A09005603, No. 2016R1D1A1B02012900; Radiation Science Research Institute, Foreign Large-size Research Facility Application Supporting project and the Global Science Experimental Data Hub Center of the Korea Institute of Science and Technology Information;
the Foundation for Fundamental Research on Matter of the Netherlands;
the Research Council of Norway;
the Polish Ministry of Science and Higher Education and 
the National Science Center;
the Ministry of Education and Science of the Russian Federation and
the Russian Foundation for Basic Research;
the Slovenian Research Agency;
Ikerbasque, Basque Foundation for Science and
MINECO (Juan de la Cierva), Spain;
the Swiss National Science Foundation; 
the Ministry of Education and the Ministry of Science and Technology of Taiwan;
the Science and Technology Facilities Council of the United Kingdom;
the Binational Science Foundation (U.S.-Israel);
and the U.S.\ Department of Energy and the National Science Foundation.
Individuals have received support from the Marie Curie EIF (European Union) and the A.~P.~Sloan Foundation (USA).

\end{document}

%% file: common_definitions.tex
\def\MDzero{\ensuremath{M_{\Dz}}\xspace}
\def\deltaMDstarDzero{\ensuremath{{\rm \Delta}M}\xspace}

\def\piplussoft {\ensuremath{\pip_{s}}\xspace}

\def\MsquaredKSpiRS{\ensuremath{M_{ \KS \pi^{-} }^{2}}\xspace}
\def\MsquaredKSpiWS{\ensuremath{M_{ \KS \pi^{+} }^{2}}\xspace}
\def\Msquaredpipi{\ensuremath{M_{ \pi^{+} \pi^{-} }^{2}}\xspace}

\def\Mbcprime{\ensuremath{M_{\mathrm{bc}}^{\prime}}\xspace}
\def\DeltaE{\ensuremath{\Delta E}\xspace}
\def\NNout{\ensuremath{\mathcal{C}_{N\kern-0.2emN_{\mathrm{out}}}}\xspace}
\def\NNoutmin{\ensuremath{\mathcal{C}^{\mathrm{min}}_{N\kern-0.2emN_{\mathrm{out}}}}\xspace}
\def\NNoutmax{\ensuremath{\mathcal{C}^{\mathrm{max}}_{N\kern-0.2emN_{\mathrm{out}}}}\xspace}
\def\NNoutprime{\ensuremath{\mathcal{C}_{N\kern-0.2emN_{\mathrm{out}}}^{\prime}}\xspace}

\def\Deltat{\ensuremath{\Delta t}\xspace}

\def\tauBzero{\ensuremath{\tau_{\Bz}}\xspace}

\def\tauBplus{\ensuremath{\tau_{\Bp}}\xspace}

\def\Deltam{\ensuremath{\Delta m_{d}}\xspace}

\def\DstarplustoDzeropisoft {\ensuremath{ \Dstarp \to \Dz \piplussoft}\xspace}

\def\DzerotoKSpipi {\ensuremath{\Dz \to \KS \pip \pim}\xspace}

\def\DtoKSpipi {\ensuremath{\D \to \KS \pip \pim}\xspace}

\def\BtoDhzero {\ensuremath{\Bz \to \D^{(*)} h^{0}}\xspace}

\newcommand*{\sPlot}{\ensuremath{{}_s{\mathcal P}\mkern-2mu\mathit{lot}}\xspace}

%% file: BABAR_BELLE_PRD_merged_author_list_final.tex
\newcommand{\TAGbbr}{${^\mathrm{A}}$}
\newcommand{\TAGbel}{${^\mathrm{B}}$}
\noaffiliation

\affiliation{Laboratoire d'Annecy-le-Vieux de Physique des Particules (LAPP), Universit\'e de Savoie, CNRS/IN2P3,  F-74941 Annecy-Le-Vieux, France}

\affiliation{Universitat de Barcelona, Facultat de Fisica, Departament ECM, E-08028 Barcelona, Spain }

\affiliation{INFN Sezione di Bari and Dipartimento di Fisica, Universit\`a di Bari, I-70126 Bari, Italy }

\affiliation{University of the Basque Country UPV/EHU, 48080 Bilbao, Spain }

\affiliation{Beihang University, Beijing 100191, China }

\affiliation{University of Bergen, Institute of Physics, N-5007 Bergen, Norway }

\affiliation{Lawrence Berkeley National Laboratory and University of California, Berkeley, California 94720, USA }

\affiliation{Ruhr Universit\"at Bochum, Institut f\"ur Experimentalphysik 1, D-44780 Bochum, Germany }

\affiliation{Institute of Particle Physics$^{\,a}$; University of British Columbia$^{b}$, Vancouver, British Columbia, Canada V6T 1Z1 }

\affiliation{Brookhaven National Laboratory, Upton, New York 11973, USA }

\affiliation{Budker Institute of Nuclear Physics SB RAS, Novosibirsk 630090, Russian Federation }

\affiliation{Novosibirsk State University, Novosibirsk 630090, Russian Federation }

\affiliation{Novosibirsk State Technical University, Novosibirsk 630092, Russian Federation }

\affiliation{University of California at Irvine, Irvine, California 92697, USA }

\affiliation{University of California at Riverside, Riverside, California 92521, USA }

\affiliation{University of California at Santa Cruz, Institute for Particle Physics, Santa Cruz, California 95064, USA }

\affiliation{California Institute of Technology, Pasadena, California 91125, USA }

\affiliation{Faculty of Mathematics and Physics, Charles University, 121 16 Prague, Czech Republic }

\affiliation{Chonnam National University, Kwangju 660-701, South Korea }

\affiliation{University of Cincinnati, Cincinnati, Ohio 45221, USA }

\affiliation{University of Colorado, Boulder, Colorado 80309, USA }

\affiliation{Deutsches Elektronen--Synchrotron, 22607 Hamburg, Germany }

\affiliation{Key Laboratory of Nuclear Physics and Ion-beam Application (MOE) and Institute of Modern Physics, Fudan University, Shanghai 200443, China }

\affiliation{INFN Sezione di Ferrara$^{a}$; Dipartimento di Fisica e Scienze della Terra, Universit\`a di Ferrara$^{b}$, I-44122 Ferrara, Italy }

\affiliation{INFN Laboratori Nazionali di Frascati, I-00044 Frascati, Italy }

\affiliation{INFN Sezione di Genova, I-16146 Genova, Italy }

\affiliation{Justus-Liebig-Universit\"at Gie\ss{}en, 35392 Gie\ss{}en, Germany }

\affiliation{SOKENDAI (The Graduate University for Advanced Studies), Hayama 240-0193, Japan }

\affiliation{Hanyang University, Seoul 133-791, South Korea }

\affiliation{University of Hawaii, Honolulu, Hawaii 96822, USA }

\affiliation{High Energy Accelerator Research Organization (KEK), Tsukuba 305-0801, Japan }

\affiliation{J-PARC Branch, KEK Theory Center, High Energy Accelerator Research Organization (KEK), Tsukuba 305-0801, Japan }

\affiliation{Humboldt-Universit\"at zu Berlin, Institut f\"ur Physik, D-12489 Berlin, Germany }

\affiliation{IKERBASQUE, Basque Foundation for Science, 48013 Bilbao, Spain }

\affiliation{Indian Institute of Science Education and Research Mohali, SAS Nagar, 140306, India }

\affiliation{Indian Institute of Technology Bhubaneswar, Satya Nagar 751007, India }

\affiliation{Indian Institute of Technology Guwahati, Assam 781039, India }

\affiliation{Indian Institute of Technology Hyderabad, Telangana 502285, India }

\affiliation{Indian Institute of Technology Madras, Chennai 600036, India }

\affiliation{Indiana University, Bloomington, Indiana 47408, USA }

\affiliation{Institute of High Energy Physics, Chinese Academy of Sciences, Beijing 100049, China }

\affiliation{Institute of High Energy Physics, Vienna 1050, Austria }

\affiliation{University of Iowa, Iowa City, Iowa 52242, USA }

\affiliation{Iowa State University, Ames, Iowa 50011, USA }

\affiliation{Advanced Science Research Center, Japan Atomic Energy Agency, Naka 319-1195}

\affiliation{J. Stefan Institute, 1000 Ljubljana, Slovenia }

\affiliation{Johns Hopkins University, Baltimore, Maryland 21218, USA }

%\affiliation{Kanagawa University, Yokohama 221-8686, Japan }

\affiliation{Institut f\"ur Experimentelle Teilchenphysik, Karlsruher Institut f\"ur Technologie, 76131 Karlsruhe, Germany }

\affiliation{Kennesaw State University, Kennesaw, Georgia 30144, USA }

\affiliation{King Abdulaziz City for Science and Technology, Riyadh 11442, Kingdom of Saudi Arabia }

\affiliation{Department of Physics, Faculty of Science, King Abdulaziz University, Jeddah 21589, Kingdom of Saudi Arabia }

\affiliation{Korea Institute of Science and Technology Information, Daejeon 305-806, South Korea }

\affiliation{Korea University, Seoul 136-713, South Korea }

\affiliation{Kyungpook National University, Daegu 702-701, South Korea }

\affiliation{Laboratoire de l'Acc\'el\'erateur Lin\'eaire, IN2P3/CNRS et Universit\'e Paris-Sud 11, Centre Scientifique d'Orsay, F-91898 Orsay Cedex, France }

\affiliation{\'Ecole Polytechnique F\'ed\'erale de Lausanne (EPFL), Lausanne 1015, Switzerland }

\affiliation{Lawrence Livermore National Laboratory, Livermore, California 94550, USA }

\affiliation{P.N. Lebedev Physical Institute of the Russian Academy of Sciences, Moscow 119991, Russian Federation }

\affiliation{Laboratoire Leprince-Ringuet, Ecole Polytechnique, CNRS/IN2P3, F-91128 Palaiseau, France }

\affiliation{University of Liverpool, Liverpool L69 7ZE, United Kingdom }

\affiliation{Faculty of Mathematics and Physics, University of Ljubljana, 1000 Ljubljana, Slovenia }

\affiliation{Queen Mary, University of London, London, E1 4NS, United Kingdom }

\affiliation{University of London, Royal Holloway and Bedford New College, Egham, Surrey TW20 0EX, United Kingdom }

\affiliation{University of Louisville, Louisville, Kentucky 40292, USA }

\affiliation{Ludwig Maximilians University, 80539 Munich, Germany }

\affiliation{Luther College, Decorah, Iowa 52101, USA }

\affiliation{Johannes Gutenberg-Universit\"at Mainz, Institut f\"ur Kernphysik, D-55099 Mainz, Germany }

%\affiliation{University of Malaya, 50603 Kuala Lumpur, Malaysia }

\affiliation{University of Manchester, Manchester M13 9PL, United Kingdom }

\affiliation{University of Maribor, 2000 Maribor, Slovenia }

\affiliation{University of Maryland, College Park, Maryland 20742, USA }

\affiliation{Massachusetts Institute of Technology, Laboratory for Nuclear Science, Cambridge, Massachusetts 02139, USA }

\affiliation{Max-Planck-Institut f\"ur Physik, 80805 M\"unchen, Germany }

\affiliation{Institute of Particle Physics$^{\,a}$; McGill University$^{b}$, Montr\'eal, Qu\'ebec, Canada H3A 2T8 }

\affiliation{School of Physics, University of Melbourne, Victoria 3010, Australia }

\affiliation{INFN Sezione di Milano$^{a}$; Dipartimento di Fisica, Universit\`a di Milano$^{b}$, I-20133 Milano, Italy }

\affiliation{University of Mississippi, University, Mississippi 38677, USA }

\affiliation{University of Miyazaki, Miyazaki 889-2192, Japan }

\affiliation{Universit\'e de Montr\'eal, Physique des Particules, Montr\'eal, Qu\'ebec, Canada H3C 3J7  }

\affiliation{Moscow Physical Engineering Institute, Moscow 115409, Russian Federation }

\affiliation{Moscow Institute of Physics and Technology, Moscow Region 141700, Russian Federation }

\affiliation{Graduate School of Science, Nagoya University, Nagoya 464-8602, Japan }

\affiliation{INFN Sezione di Napoli$^{a}$ and Dipartimento di Scienze Fisiche, Universit\`a di Napoli Federico II$^{b}$, I-80126 Napoli, Italy }

\affiliation{Nara Women's University, Nara 630-8506, Japan }

\affiliation{National Central University, Chung-li 32054, Taiwan }

\affiliation{National United University, Miao Li 36003, Taiwan }

\affiliation{NIKHEF, National Institute for Nuclear Physics and High Energy Physics, NL-1009 DB Amsterdam, The Netherlands }

\affiliation{Department of Physics, National Taiwan University, Taipei 10617, Taiwan }

\affiliation{H. Niewodniczanski Institute of Nuclear Physics, Krakow 31-342, Poland }

\affiliation{Nippon Dental University, Niigata 951-8580, Japan }

\affiliation{Niigata University, Niigata 950-2181, Japan }

\affiliation{University of Notre Dame, Notre Dame, Indiana 46556, USA }

\affiliation{Ohio State University, Columbus, Ohio 43210, USA }

\affiliation{Novosibirsk State University, Novosibirsk 630090, Russian Federation }

\affiliation{Osaka City University, Osaka 558-8585, Japan }

\affiliation{Pacific Northwest National Laboratory, Richland, Washington 99352, USA }

\affiliation{INFN Sezione di Padova$^{a}$; Dipartimento di Fisica, Universit\`a di Padova$^{b}$, I-35131 Padova, Italy }

\affiliation{Panjab University, Chandigarh 160014, India }

\affiliation{Laboratoire de Physique Nucl\'eaire et de Hautes Energies, IN2P3/CNRS, Universit\'e Pierre et Marie Curie-Paris6, Universit\'e Denis Diderot-Paris7, F-75252 Paris, France }

\affiliation{Peking University, Beijing 100871, China }

\affiliation{INFN Sezione di Perugia$^{a}$; Dipartimento di Fisica, Universit\`a di Perugia$^{b}$, I-06123 Perugia, Italy }

\affiliation{INFN Sezione di Pisa$^{a}$; Dipartimento di Fisica, Universit\`a di Pisa$^{b}$; Scuola Normale Superiore di Pisa$^{c}$, I-56127 Pisa, Italy }

\affiliation{University of Pittsburgh, Pittsburgh, Pennsylvania 15260, USA }

\affiliation{Princeton University, Princeton, New Jersey 08544, USA }

\affiliation{Theoretical Research Division, Nishina Center, RIKEN, Saitama 351-0198, Japan }

\affiliation{INFN Sezione di Roma$^{a}$; Dipartimento di Fisica, Universit\`a di Roma La Sapienza$^{b}$, I-00185 Roma, Italy }

\affiliation{Universit\"at Rostock, D-18051 Rostock, Germany }

\affiliation{Rutherford Appleton Laboratory, Chilton, Didcot, Oxon, OX11 0QX, United Kingdom }

\affiliation{CEA, Irfu, SPP, Centre de Saclay, F-91191 Gif-sur-Yvette, France }

\affiliation{University of Science and Technology of China, Hefei 230026, China }

\affiliation{Showa Pharmaceutical University, Tokyo 194-8543, Japan }

\affiliation{Soongsil University, Seoul 156-743, South Korea }

\affiliation{SLAC National Accelerator Laboratory, Stanford, California 94309 USA }

\affiliation{University of South Carolina, Columbia, South Carolina 29208, USA }

\affiliation{University of South Carolina, Columbia, South Carolina 29208, USA }

\affiliation{Southern Methodist University, Dallas, Texas 75275, USA }

\affiliation{St. Francis Xavier University, Antigonish, Nova Scotia, Canada B2G 2W5 }

\affiliation{Stanford University, Stanford, California 94305, USA }

\affiliation{State University of New York, Albany, New York 12222, USA }

\affiliation{Sungkyunkwan University, Suwon 440-746, South Korea }

\affiliation{School of Physics, University of Sydney, New South Wales 2006, Australia }

\affiliation{Department of Physics, Faculty of Science, University of Tabuk, Tabuk 71451, Kingdom of Saudi Arabia }

\affiliation{Tata Institute of Fundamental Research, Mumbai 400005, India }

\affiliation{Department of Physics, Technische Universit\"at M\"unchen, 85748 Garching, Germany }

\affiliation{Excellence Cluster Universe, Technische Universit\"at M\"unchen, 85748 Garching, Germany }

\affiliation{Tel Aviv University, School of Physics and Astronomy, Tel Aviv, 69978, Israel }

\affiliation{University of Tennessee, Knoxville, Tennessee 37996, USA }

\affiliation{University of Texas at Austin, Austin, Texas 78712, USA }

\affiliation{University of Texas at Dallas, Richardson, Texas 75083, USA }

\affiliation{Toho University, Funabashi 274-8510, Japan }

\affiliation{Department of Physics, Tohoku University, Sendai 980-8578, Japan }

\affiliation{Earthquake Research Institute, University of Tokyo, Tokyo 113-0032, Japan }

\affiliation{Department of Physics, University of Tokyo, Tokyo 113-0033, Japan }

\affiliation{Tokyo Institute of Technology, Tokyo 152-8550, Japan }

\affiliation{Tokyo Metropolitan University, Tokyo 192-0397, Japan }

\affiliation{INFN Sezione di Torino$^{a}$; Dipartimento di Fisica, Universit\`a di Torino$^{b}$, I-10125 Torino, Italy }

\affiliation{INFN Sezione di Trieste and Dipartimento di Fisica, Universit\`a di Trieste, I-34127 Trieste, Italy }

\affiliation{IFIC, Universitat de Valencia-CSIC, E-46071 Valencia, Spain }

\affiliation{Institute of Particle Physics$^{\,a}$; University of Victoria$^{b}$, Victoria, British Columbia, Canada V8W 3P6 }

\affiliation{Virginia Polytechnic Institute and State University, Blacksburg, Virginia 24061, USA }

\affiliation{Department of Physics, University of Warwick, Coventry CV4 7AL, United Kingdom }

\affiliation{Wayne State University, Detroit, Michigan 48202, USA }

\affiliation{University of Wisconsin, Madison, Wisconsin 53706, USA }

\affiliation{Yamagata University, Yamagata 990-8560, Japan }

\affiliation{Yonsei University, Seoul 120-749, South Korea }

\author{I.~Adachi\TAGbel}\affiliation{High Energy Accelerator Research Organization (KEK), Tsukuba 305-0801, Japan }\affiliation{SOKENDAI (The Graduate University for Advanced Studies), Hayama 240-0193, Japan } % KEK
\author{T.~Adye\TAGbbr}\affiliation{Rutherford Appleton Laboratory, Chilton, Didcot, Oxon, OX11 0QX, United Kingdom }
\author{H.~Ahmed\TAGbbr}\affiliation{St. Francis Xavier University, Antigonish, Nova Scotia, Canada B2G 2W5 }
\author{J.~K.~Ahn\TAGbel}\affiliation{Korea University, Seoul 136-713, South Korea } % Korea
\author{H.~Aihara\TAGbel}\affiliation{Department of Physics, University of Tokyo, Tokyo 113-0033, Japan } % Tokyo
\author{S.~Akar\TAGbbr}\affiliation{Laboratoire de Physique Nucl\'eaire et de Hautes Energies, IN2P3/CNRS, Universit\'e Pierre et Marie Curie-Paris6, Universit\'e Denis Diderot-Paris7, F-75252 Paris, France }
\author{M.~S.~Alam\TAGbbr}\affiliation{State University of New York, Albany, New York 12222, USA }
\author{J.~Albert\TAGbbr$^{b}$}\affiliation{Institute of Particle Physics$^{\,a}$; University of Victoria$^{b}$, Victoria, British Columbia, Canada V8W 3P6 }
\author{F.~Anulli\TAGbbr$^{a}$}\affiliation{INFN Sezione di Roma$^{a}$; Dipartimento di Fisica, Universit\`a di Roma La Sapienza$^{b}$, I-00185 Roma, Italy }
\author{N.~Arnaud\TAGbbr}\affiliation{Laboratoire de l'Acc\'el\'erateur Lin\'eaire, IN2P3/CNRS et Universit\'e Paris-Sud 11, Centre Scientifique d'Orsay, F-91898 Orsay Cedex, France }
% ...
% Ljubljana

\collaboration{The {\TAGbbr}\babar\ and {\TAGbel}Belle Collaborations}

%% file: PRD_paper_B0ToD0h0_DToK0Spipi.bbl
\begin{thebibliography}{99}
\bibitem{phi_three} The CKM angle $\gamma$ is also referred to as $\phi_3$ in the literature.
\end{thebibliography}
